\documentclass{aa}
\usepackage{graphicx} 
\usepackage[varg]{txfonts}
\usepackage{orcidlink}
\usepackage{placeins}
\usepackage{hyperref}
\usepackage{longtable}
\usepackage{multirow}
\hypersetup{
    colorlinks=true,
    linkcolor=blue,
    filecolor=magenta,      
    urlcolor=blue,
    citecolor=blue,
    pdftitle={Overleaf Example},
    }
\usepackage{afterpage}

\authorrunning{Zeng et al.}
\titlerunning{sGRB energetics and Remnant BH magnetism}

\begin{document}

\title{
Jet energetics and remnant magnetism in short gamma-ray bursts from binary neutron-star mergers: A statistical study via broadband afterglow modeling
}

   \author{Changwen Zeng\orcidlink{0009-0006-3194-8525}  \inst{1,2}\fnmsep
          \and
          Zhiwei Chen\orcidlink{0000-0001-7952-7945}\inst{1,3}\thanks{chenzhiwei171@mails.ucas.ac.cn}
          \and
         Youjun Lu\orcidlink{0000-0002-1310-4664}\inst{2,1}\thanks{luyj@nao.cas.cn}
         \and
         Hui Sun\orcidlink{0000-0002-9615-1481}
         \inst{1,2}
          \and
          Jihui Zhang\orcidlink{0009-0001-7743-2747}\inst{1,2}}

   \institute{National Astronomical Observatories, Chinese Academy of Sciences, 20A Datun Road, Beijing 100101, China 
    \and
    School of Astronomy and Space Sciences, University of Chinese Academy of Sciences, 19A Yuquan Road, Beijing 100049, China
    \and Institut f\"ur Theoretische Physik, Goethe Universit\"at, Max-von-Laue-Str. 1, 60438 Frankfurt am Main, Germany
    }

\date{}
\abstract{
\textit{Context}: Short gamma-ray bursts (sGRBs) are widely believed to be associated with compact binary mergers, especially binary neutron star (BNS) mergers, and thus provide a probe of relativistic jet energetics and merger remnant properties. \\[\medskipamount]
\textit{Aims}: We aim to derive statistical constraints on the characteristic energetics of sGRBs and to explore the implications of these constraints on the magnetic field strength of the remnant black holes (BHs) required to power relativistic jets. \\[\medskipamount]
\textit{Methods}: We compile a broadband afterglow dataset of 40 sGRBs and perform burst-by-burst forward shock modeling using an off-axis Gaussian jet framework after removing the data potentially affected by additional energy injection. We derive the statistical distributions of the jet kinetic energy $E_{\rm k}$ and radiative efficiency $\eta_{\gamma}$. With the above distributions and results of compact binary population synthesis and numerical simulations, we estimate the characteristic magnetic field strength of BNS merger remnant BHs under the assumption of Blandford–Znajek (BZ) mechanism for jet launching.\\[\medskipamount]
\textit{Results}:{ 
For the 15 strictly constrained bursts, we infer a median beaming-corrected kinetic energy of $E_\mathrm{k}\simeq 3.8\times10^{49}$~erg and a radiative efficiency of $\eta_\gamma\simeq0.35$, with corresponding $68\%$ intervals of $[5.1\times10^{48},1.9\times10^{50}]$~erg and [0.09,0.79], respectively. Assuming BZ-powered jets, we infer a characteristic magnetic field supplied by the surrounding accretion disk and threading the BH horizon of approximately $7\times10^{15}$~G, with a $68\%$ credible interval of approximately $[2.3\times10^{15},2.4\times10^{16}]$~G, depending slightly on the assumed neutron star equation of state. Extending the analysis to the full sample yields a higher median field strength of approximately $2\times10^{16}$~G, with the high-energy and high-field tails predominantly contributed by less well-constrained bursts.
}
\\[\medskipamount] 
\textit{Conclusions}:
Our results highlight two key implications. First, statistical modeling of the broadband afterglow emission of sGRBs within a structured jet framework provides robust constraints on the jet energetics. Second, the inferred energetics enable us to probe the properties of the central engine, and imply a large BH magnetic field strength on the order of $B_\mathrm{BH}\sim10^{16}$~G. Comparing these values with the magnetic field distribution of pre-merger NSs suggests a characteristic magnetic field amplification factor of $\eta_B\sim10^3$. Such a large amplification factor provides an independent observational constraint on magnetic field amplification processes operating during BNS mergers and the subsequent formation of the BH system.
}

\keywords{ gamma-ray burst: general -- stars: neutron -- black hole physics -- magnetic fields -- methods: statistical
}

\maketitle

\section{Introduction}
\label{sec:intro}

Gamma-ray bursts (GRBs) are among the most luminous transient phenomena in the Universe and are widely regarded as promising electromagnetic counterparts to gravitational wave (GW) sources. Short GRBs (sGRBs), defined by prompt emission durations $T_{90}<2$~s \citep{kmf+93}, are widely believed to be linked to compact binary mergers. The landmark detection of the binary neutron star (BNS) merger GW170817, together with its associated sGRB GRB~170817A, provided the first direct and compelling evidence that sGRBs can originate from BNS mergers \citep{aaa+17,2017ApJ...848L..14G,2017ApJ...848L..15S}. This discovery established a firm physical connection between sGRBs and BNS mergers and marked the beginning of a new era of multi-messenger astronomy.

Multi-wavelength observations of sGRBs, in particular their broadband afterglow emission, provide a powerful probe of the properties of relativistic jets launched during these merger events. The afterglow emission arises from the interaction between the relativistic jet and the surrounding circumburst medium \citep{1997ApJ...482L..29M,1998ApJ...497L..17S,1999ApJ...523..177W}. By modeling the broadband afterglow light curves, one could constrain key physical properties such as the jet geometry, energy, and ambient density, thereby gaining insight into the launching and evolution of ultra-relativistic outflows \citep[e.g.,][]{rgb+11,2015ApJ...806L..14P,2016ApJ...824L...6R,otd+21,2024A&A...692A..73G}. 

Accurate inference of jet energetics, however, depends on adopting a realistic description of the jet structure. In recent years, growing observational evidence has favored structured off-axis jet models over simple on-axis top-hat configurations \citep{ccr+20,lhw+22}. In the off-axis model, the energy per unit solid angle varies smoothly with polar angle from the jet axis, producing characteristic afterglow signatures for off-axis observers, such as delayed peaks and complex light curve evolution \citep{rvp+20,lga+23,rwd+24}. A prominent example is GRB~170817A, whose afterglow exhibited a slow rise followed by a sharp decline, naturally explained by an off-axis structured jet \citep[e.g.,][]{2017MNRAS.472.4953L,2018ApJ...863L..18A,2018ApJ...869...55W,2018Natur.561..355M,2018RAA....18...43H,2019ApJ...883L...1F,2019ApJ...870L..15L}. 

Beyond constraining jet geometry, broadband afterglow modeling enables robust estimates of the beaming-corrected kinetic energy and thus the total energetics of sGRBs. The energy distribution of sGRBs is a key ingredient for understanding their detection rates and observational selection effects \citep{wp+15,gsp+16,aaaa+17}. More importantly, the energetics of sGRBs provide an observational link to the physical properties of the central engine formed after the BNS merger. In particular, if the relativistic jet is powered by the Blandford--Znajek (BZ) mechanism, the jet energy can be directly related to the properties of the merger remnant BH. For example, \citet{ll+24} investigated the mass and spin evolution of BHs formed in BNS mergers using inferred jet luminosities, while \citet{lfm+19} proposed a framework linking GRB energetics to the magnetic field strength of the central BH. However, these studies often rely on limited samples, simplified assumptions, or energetics that are not uniformly constrained by broadband afterglow modeling. In this work, we compile broadband afterglow observations of 40 sGRBs and analyze them within a uniform modeling framework, thereby constructing, to the best of our knowledge, the largest homogeneously analyzed broadband afterglow sample of sGRBs to date. From these fits, we derive the distribution of jet energetics of sGRBs. Subsequently, by assuming that the jets are powered by the BZ process, we then combine the inferred jet energetics distribution with BNS merger properties predicted by binary population synthesis calculations. The remnant BH properties are then obtained using fitting formulae calibrated from numerical relativity (NR) simulations, allowing us to estimate the magnetic field strengths of the newly formed BHs.

This paper is organized as follows. In Section~\ref{sec:sample}, we introduce the sGRB sample used in this work. Section~\ref{sec:analysis} describes the processing of the broadband afterglow data, including the homogenization of multi-wavelength observations and the burst-by-burst selection of data dominated by forward shock (FS) emission. In Section~\ref{sec:fitting}, we present the FS modeling based on an off-axis Gaussian structured jet, together with the fitting procedure and the resulting distributions of GRB energetics. In Section~\ref{sec:BZ}, we combine the inferred energetic distributions with a synthetic compact binary population model to estimate the BH magnetic field strength within the framework of the BZ process, and discuss the associated physical implications. The impact of the main assumptions adopted in our analysis is examined in Section~\ref{sec:discussion}. Finally, our conclusions are summarized in Section~\ref{sec:conclusion}.

\section{The sample}
\label{sec:sample}

The GRBs considered in this work are drawn from the catalog presented by \citet{nfd+22}, which was constructed by identifying GRBs that are likely associated with compact binary mergers, particularly BNS systems. It primarily consists of bursts with $T_{90} < 2$\,s, which are commonly interpreted as originating from compact object mergers, based on both their short duration and their statistical association with older host galaxy populations. In addition, the sample includes events with well-localized afterglows and a small number of long-duration GRBs that exhibit observational signatures consistent with BNS mergers, such as kilonova emission. These observational properties collectively make this sample well suited for studies of the physical properties of merger remnants. In their study, the authors analyzed 84 GRB host galaxies with broadband photometric detections and upper limits, and identified host associations using the probability of chance coincidence ($P_{\rm cc}$). This method compares the optical magnitude of a candidate host and its angular offset from the GRB position (as determined from X-ray, optical, or radio afterglow observations) to estimate the likelihood of a physical association. They further defined a ``Gold Sample'' consisting of events with robust host identifications ($P_{\rm cc} < 0.02$). These GRBs have well-established redshifts, which are essential for accurately determining the energetics of the bursts in our modeling. 
For example, GRB~230307A exhibits a well-sampled multi-wavelength afterglow and an associated kilonova. However, its proposed host galaxy at $z\simeq0.065$ has a relatively high chance-coincidence probability of $P_{\rm cc}\sim0.09$ \citep{2024Natur.626..737L}, which does not satisfy the $P_{\rm cc}<0.02$ criterion adopted for our sample. We therefore do not include GRB~230307A in the present analysis.
A similar analysis was performed by \citet{ji+24}, who identified additional bursts consistent with the same selection criteria. We adopt the same $P_{\mathrm{cc}}$ threshold and select four additional bursts (GRB 060505, 060801, 210510A, and 211227A). Among them, the origin of GRB 060505 is still under debate \citep{tfo+08,jzc+21}, so we only include the latter three bursts in our final sample, thereby minimizing possible contamination from events with uncertain progenitor origins.

For the following analysis, we collect the available radio-to-X-ray afterglow data for the Gold Sample. The procedures for data selection and reduction in each band are introduced below. The redshifts and the 15–150\,keV fluences of the bursts from Swift Burst Alert Telescope (BAT) analysis\footnote{\href{https://www.swift.ac.uk/analysis/bat/}{https://www.swift.ac.uk/analysis/bat/}} are listed in Table~\ref{tab:redshift}.

\begin{table}
\caption{Short GRB redshifts and 15--150 keV fluences.}
\label{tab:redshift}
\centering
\begin{tabular}{lccc}
\hline\hline
\multirow{2}{*}{GRB} & \multirow{2}{*}{Redshift} & 15--150 keV fluence & Error \\
    &          & (erg cm$^{-2}$)     & (erg cm$^{-2}$) \\
\hline
211227A & 0.228 & $8.10\times10^{-6}$ & $2.3\times10^{-7}$ \\
211211A & 0.076 & $1.46\times10^{-4}$ & $2\times10^{-6}$ \\
211023B & 0.862 & $1.68\times10^{-7}$ & $2.7\times10^{-8}$ \\
210726A & 2.38 & $4.32\times10^{-8}$ & $1.06\times10^{-8}$ \\
210323A & 0.733 & $2.44\times10^{-7}$ & $2.6\times10^{-8}$ \\
200907B & 0.56 & $1.60\times10^{-7}$ & $2.0\times10^{-8}$ \\
200522A & 0.554 & $1.11\times10^{-7}$ & $1.3\times10^{-8}$ \\
200219A & 0.48 & $1.30\times10^{-6}$ & $2.0\times10^{-7}$ \\
181123B & 1.754 & $1.27\times10^{-7}$ & $2.0\times10^{-8}$ \\
180727A & 1.95 & $2.92\times10^{-7}$ & $2.0\times10^{-8}$ \\
180418A & 1.55 & $3.42\times10^{-7}$ & $3.3\times10^{-8}$ \\
170817A & 0.01 & -\tablefootmark{a} & -\tablefootmark{a} \\
170428A & 0.453 & $2.80\times10^{-7}$ & $2.2\times10^{-8}$ \\
160525B & 0.64 & $3.14\times10^{-8}$ & $7.6\times10^{-9}$ \\
160411A & 0.81 & $7.84\times10^{-8}$ & $1.81\times10^{-8}$ \\
150728A & 0.461 & $3.26\times10^{-8}$ & $9.7\times10^{-9}$ \\
150120A & 0.46 & $1.44\times10^{-7}$ & $1.6\times10^{-8}$ \\
150101B & 0.134 & $6.10\times10^{-8}$ & $2.20\times10^{-8}$ \\
141212A & 0.596 & $7.24\times10^{-8}$ & $1.16\times10^{-8}$ \\
140903A & 0.351 & $1.35\times10^{-7}$ & $1.0\times10^{-8}$ \\
140129B & 0.43 & $7.08\times10^{-8}$ & $1.41\times10^{-8}$ \\
130603B & 0.357 & $6.27\times10^{-7}$ & $2.6\times10^{-8}$ \\
121226A & 1.37 & $1.42\times10^{-7}$ & $2.4\times10^{-8}$ \\
120804A & 1.05 & $8.78\times10^{-7}$ & $4.5\times10^{-8}$ \\
101224A & 0.454 & $5.63\times10^{-8}$ & $1.08\times10^{-8}$ \\
100206A & 0.407 & $1.39\times10^{-7}$ & $5.1\times10^{-8}$ \\
100117A & 0.914 & $9.35\times10^{-8}$ & $1.27\times10^{-8}$ \\
090510 & 0.903 & $1.03\times10^{-7}$ & $4.3\times10^{-8}$ \\
071227 & 0.381 & $4.94\times10^{-7}$ & $1.22\times10^{-7}$ \\
070809 & 0.473 & $1.02\times10^{-7}$ & $1.5\times10^{-8}$ \\
070724A & 0.457 & $3.10\times10^{-8}$ & $6.8\times10^{-9}$ \\
070714B & 0.923 & $7.39\times10^{-7}$ & $9.3\times10^{-8}$ \\
070429B & 0.902 & $6.52\times10^{-8}$ & $1.00\times10^{-8}$ \\
061210 & 0.41 & $1.09\times10^{-6}$ & $1.9\times10^{-7}$ \\
061006 & 0.461 & $1.43\times10^{-6}$ & $1.4\times10^{-7}$ \\
060614 & 0.125 & $2.08\times10^{-5}$ & $3.00\times10^{-7}$ \\
051221A & 0.546 & $1.16\times10^{-6}$ & $4\times10^{-8}$ \\
050724 & 0.257 & $1.00\times10^{-6}$ & $1.2\times10^{-7}$ \\
050709 & 0.161 & -\tablefootmark{a} & -\tablefootmark{a} \\
050509B & 0.225 & $7.13\times10^{-9}$ & $2.04\times10^{-9}$ \\
\hline
\end{tabular}
\tablefoot{
Redshifts for all bursts are taken from \citet{fnd+22}, with the exception of GRB 210726A. Its redshift has been revisited and clarified in \citet{srl+24}.\\
\tablefoottext{a}{ These two bursts did not trigger Swift BAT.
}
}
\end{table}
\section{Data processing}
\label{sec:analysis}

In this section, we present the broadband observations of the Gold Sample and compile the afterglow light curve data after excluding components that are not associated with FS emission. Section~\ref{sub:broad} describes the reduction of the broadband observational data. In Section~\ref{sub:fs}, we outline the criteria used to identify FS-dominated emission for each burst. The resulting FS-dominated X-ray, optical/near-infrared, and radio afterglow data are summarized in Tables~\ref{tab:xdata}, \ref{tab:otable}, and \ref{tab:rtable}.

\subsection{Broadband data}
\label{sub:broad}

For each burst, we collect data of X-ray afterglow observations from the literature when available; otherwise, we take the data given by the \textit{Swift}/X-ray Telescope (XRT) observations \citep{bhn+05} of GRB afterglows. XRT operates in the 0.3–10\,keV band with an effective area of about 120\,$\mathrm{cm^{2}}$  at $1.5$\,keV and an angular resolution of 18\,arcsec (half-power diameter). XRT can autonomously slew to GRB positions within approximately 100 seconds after the trigger, providing early-time X-ray light curves and spectra crucial for constraining the afterglow physics. For the \textit{Swift} X-ray afterglow data, when available, we adopt the unabsorbed flux density light curves directly from the \textit{Swift} XRT repository \citep{ebp+07, ebp+09}. In cases where only count rate light curves are provided, we convert them to unabsorbed fluxes using the appropriate counts-to-flux conversion factor, and derive the flux densities using the photon index $\Gamma$, assuming a power-law spectrum of the form $F_{\nu, \mathrm{X}} \propto \nu^{1 - \Gamma}$.

For the optical/near-infrared (NIR) afterglow data, we compile available observations reported in the GCN Circulars\footnote{\href{https://gcn.gsfc.nasa.gov/selected.html}{https://gcn.gsfc.nasa.gov/selected.html}} and in the literature for each burst. The dataset includes measurements obtained with a variety of telescopes and instruments, including the \textit{Swift} Ultra-Violet/Optical Telescope (UVOT; \citealt{rkm+05}) and the Gemini Multi-Object Spectrograph (GMOS; \citealt{hja+04}). Magnitudes are converted to the AB system using instrument-specific conversion factors, or based on \citet{will+08} when unavailable. All magnitudes are corrected for Galactic extinction along the line of sight following \citet{fit+99}, and converted to flux densities using the relation in \citet{fig+96}, with host galaxy extinction accounted for where possible.

For the radio observations, we adopt the flux densities reported in the literature directly and we therefore do not apply any additional corrections.

\subsection{Forward shock data selection}
\label{sub:fs}

In our fitting procedure, we include only the data associated with the FS emission, as this component provides a relatively clean and theoretically well-established probe of the jet kinetic energy within the standard fireball framework \citep{p+99,zm+03}. The FS emission corresponds to the broadband afterglow radiation generated as the relativistic jet decelerates in the external medium, and its dynamical evolution is directly linked to the bulk energetics of the outflow. Observations that are likely affected by reverse shock (RS) components, flares or late-time energy injection are excluded from the modeling, since these processes may introduce additional uncertainties and model dependencies related to the central engine or ejecta structure \citep[e.g.,][]{yd+07,yd+09,llt+12,gwm+15,yww+15}. In addition, when a reported detection conflicts with an upper limit, we adopt the measured flux and exclude the upper limit from the fitting. Below, we list only the bursts for which data selection is performed. For those bursts whose entire data sets are used without further selection, we do not list them individually. The resulting unabsorbed flux densities, $1\sigma$ uncertainties, and $3\sigma$ upper limits are summarized in Table~\ref{tab:xdata}, Table~\ref{tab:otable} and Table~\ref{tab:rtable}.

\begin{itemize}

\item \textit{GRB 211211A}: We include only the optical/NIR and radio data obtained within $\sim0.1$ days after the burst, as later-time emission may be contaminated by a kilonova component \citep{yaz+22}.

\item \textit{GRB 211023B}: Optical and X-ray observations prior to $\sim500$ s show an early peak inconsistent with the FS model. We attribute this feature to flare emission and exclude these data from the fitting.

\item \textit{GRB 210726A}: The early X-ray observations show a plateau followed by a sharp decay, which are unlikely to be attributed to the external shock \citep{zfd+06}. The optical upper limits may be significantly affected by host galaxy extinction, and a radio flare may also be present \citep{srl+24}. As a result, these data are excluded from our fitting.

\item \textit{GRB~210510A}: No X-ray afterglow observations are available for this burst \citep{pdd+21}. In addition, given the uncertain association of the reported optical counterpart \citep{aaa+21}, this event is excluded from our final afterglow fitting sample.

\item \textit{GRB 200522A}: The radio observation at $\sim0.1$ day may have been produced by RS emission \citep{otd+21}.

\item \textit{GRB 181123B}: Only the late-time X-ray observations are used, since the early-time excess emission is inconsistent with the standard FS model \citep{ol+18,pfn+20}.

\item \textit{GRB 180618A}: The chromatic evolution in the optical band and the steep decay of the early X-ray light curve indicate inconsistency with a standard FS scenario \citep{jmg+22}; thus, this burst is excluded from our sample.

\item \textit{GRB 180418A}: We only use the data after $\sim500$ s, since the broadband emission at earlier times is likely dominated by RS emission \citep{wzz+24}.

\item \textit{GRB 170728B}: The bright early-time optical emission \citep{dm+17} and the inconsistency between the late-time radio and X-ray data cannot be explained within the standard FS model; therefore, we exclude this burst from our sample.

\item \textit{GRB 150101B}: Since the optical emission may be contaminated by a luminous kilonova \citep{trp+18}, we only include the optical upper limits in our modeling. In addition, because no X-ray emission was detected at early times, we adopt upper limits based on the constraints at the corresponding X-ray epochs.

\item \textit{GRB 120804A}: The optical observations are excluded because they are likely heavily affected by host galaxy extinction \citep{bzl+13}.

\item \textit{GRB 071227}: An apparent optical excess may be present \citep{eow+19}, so we include only the optical upper limits. In addition, the early X-ray observations exhibit a behavior similar to that of GRB 210726A and are therefore excluded from the analysis.

\item \textit{GRB 070809}: The optical and late-time X-ray observations may originate from a kilonova \citep{jcl+20}. Therefore, we include only the optical upper limits and the early-time X-ray data in our analysis.

\item \textit{GRB 070724A}: The early X-ray observations show a steep decay, and the NIR emission likely has a non–afterglow origin \citep{bcf+09}. Therefore, we use only the late-time X-ray data and the i-band observations in our analysis.

\item \textit{GRB 070714B}: The late-time optical data may be contaminated by energy injection \citep{gzl+17}, so we take them as upper limits. The early-time X-ray observations are excluded for the same reason as in GRB 210726A.

\item \textit{GRB 070429B}: The Swift/UVOT data show a rapidly rising and then quickly fading faint afterglow \citep{kkz+11}, which is inconsistent with a standard FS model. Therefore, we include only the UVOT white-band data in our analysis.

\item \textit{GRB 061006}: The late-time optical measurements are likely contaminated by merger-nova emission \citep{gzl+17}. Therefore, we include only the upper limits in our analysis.

\item \textit{GRB 060801}: The X-ray light curve shows a behavior similar to that of GRB 070429B, and follow-up optical observations failed to detect any afterglow emission \citep{dd+09}. Thus, this burst is excluded from our sample. 

\item \textit{GRB 051221A}: Previous study has attributed the early-time X-ray emission of this object to the FS while the late-time component to a millisecond magnetar \citep{fx+06}. However, this interpretation faces several difficulties: it requires an unusually large electron energy index ($p>5$), and even models with energy injection fail to reproduce the broadband observations \citep{rvm+15}. We therefore do not attribute the early-time X-ray emission to the FS, and instead restrict our modeling to the late-time X-ray data, which are fully consistent with the standard FS scenario.

\item \textit{GRB 050724}: The X-ray and optical light curves show evidence of energy injection at $\sim10^{5}$ s \citep{mcd+07,gbp+06}, which significantly affects the afterglow behavior. Therefore, in our fitting we include only the upper limits at that epoch.

\end{itemize}

Meanwhile, for GRB 211227A, 210323A, 200219A, 180727A, 150120A, 100117A, and 060614, the early X-ray observations exhibit behavior similar to that of GRB 210726A. Therefore, we exclude these early-time data from our fitting. After applying the FS selection described above, 40 of the 45 bursts are retained in the final sample.

\section{Afterglow light curve fitting}
\label{sec:fitting}

The FS-selected data for each burst enable further light curve fitting with physically motivated afterglow models. In this work, we analyze these data within the framework of an off-axis Gaussian jet model. Section~\ref{sub:gaussian} describes the energetics of the Gaussian jet model for a non-zero viewing angle. Section~\ref{sub:afterglowpy} introduces the numerical package \texttt{afterglowpy} employed in the fitting procedure. The resulting light curves for each burst, together with the posterior distributions of the model parameters, are presented in Section~\ref{sub:results}. Finally, in Section~\ref{sub:energetics}, we characterize the distributions of several key energetics parameters using parametric models. 

\subsection{Gaussian jet energetics}
\label{sub:gaussian}

The total energy of a $\gamma$-ray jet consists of two components: the kinetic energy $E_\mathrm{k}$ and the energy radiated in $\gamma$-rays, $E_\gamma$. The prompt $\gamma$-ray efficiency is defined as
\begin{equation}
\eta_\gamma = \frac{E_{\gamma}}{E_{\gamma} + E_{\mathrm{k}}}.
\label{con:eta}
\end{equation}

The kinetic energy of the afterglow can be inferred from the observed light curves once a specific jet structure is adopted. Rather than assuming a simple top-hat jet model—where the jet energy is uniform within the opening angle and abruptly drops to zero outside—we adopt a more realistic structured jet configuration, namely a Gaussian jet profile. This choice is motivated by both numerical simulations and observations \citep{tpr+18,ktg+19,pdc+23}, which suggest a smooth angular energy structure, while also keeping the model simple and well-constrained. In this model, the kinetic energy per unit solid angle decreases smoothly with polar angle according to
\begin{equation}
4\pi\frac{\mathrm{d}E_{\rm k}}{\mathrm{d}\Omega }\equiv E(\theta) =
\begin{cases} 
E_{\mathrm{0}}\,\exp\left(-\frac{\theta^2}{2\theta_{\mathrm{c}}^2}\right),  & \mathrm{for\ \ }\theta  \leq  \theta_{\mathrm{w}}, \\
0, &  \mathrm{for\ \ } \theta  > \theta_{\mathrm{w}},
\end{cases}
\label{con:etheta}
\end{equation} 
where $E_{\mathrm{0}}$ is the on-axis kinetic energy, $\theta$ is the polar angle measured from the jet axis, $\theta_{\mathrm{c}}$ denotes the characteristic core angle, and $\theta_{\mathrm{w}}$ represents the truncation angle beyond which the energy vanishes. Integrating Equation~(\ref{con:etheta}) over the solid angle of both jets yields the beaming-corrected total kinetic energy (\citealt{ccr+20}) as
\begin{equation}
E_\mathrm{k}=E_\mathrm{0} \theta_\mathrm{c}^2\left(1-e^{-\frac{\theta_{\mathrm{w}}^2}{2\theta_{\mathrm{c}}^2}}\right).
\label{con:ek}
\end{equation} 

The radiative energy can be estimated from the prompt emission. Specifically, we estimate the isotropic-equivalent $\gamma$-ray energy $E_{\gamma,\mathrm{iso}}$ from the prompt fluence measured by the \textit{Swift}/BAT instrument in the 15--150\,keV band. 
The bolometric isotropic-equivalent $\gamma$-ray energy is given by
\begin{equation}
E_{\gamma,\mathrm{iso}} = 
k \times 
\frac{4\pi d_{\mathrm{L}}^{2}}{1+z} f_{\gamma},
%
\label{con:egammaiso}
\end{equation}
where $d_{\mathrm{L}}$ is the luminosity distance, $z$ is the redshift, $f_{\gamma}$ is the observed fluence in the 15--150 keV band, and $k$ is the bolometric $k$-correction factor, defined as
\begin{equation}
 k=\frac{\int_{1 ~\mathrm{ keV}}^{10^4 ~\mathrm{ keV}}EN(E) \mathrm{d}E}{\int_{15(1+z) ~\mathrm{ keV}}^{150(1+z) ~\mathrm{ keV}}EN(E)\mathrm{d}E }, 
\label{con:k_correction}
\end{equation}
where $N(E)$ denotes the prompt emission spectral model in the GRB rest frame. For a given observed spectral model, the peak energy is transformed as $E_{\rm peak,rest}=E_{\rm peak,obs}(1+z)$, while the spectral shape is otherwise unchanged. The adopted spectral parameters and the resulting $k$-correction factors are listed in Table~\ref{tab:k-correction}. When such spectral information is unavailable, we retain the population-averaged value $k=5$ adopted by \citet{fbm+15}. For GRB~170817A and GRB~050709, which were not detected by \textit{Swift}/BAT, we directly adopt the published bolometric $E_{\gamma,\mathrm{iso}}$ values from \citet{2018MNRAS.475.2971B} and \citet{2005Natur.437..855V}, respectively.

For a structured jet observed at a viewing angle $\theta_{\mathrm{obs}}$, the isotropic-equivalent energy depends on both the jet structure and relativistic beaming effects. Following \citet{sgp+15}, the relation between $E_{\gamma,\mathrm{iso}}$ and the true $\gamma$-ray energy $E_{\gamma}$ is given by
\begin{equation}
E_{\gamma,\mathrm{iso}}= E_{\gamma}\int_{0}^{2\pi} \mathrm{d}\phi
\int_{0}^{\pi} \mathrm{d}\theta \sin \theta 
\times \frac{\epsilon (\theta)}{\Gamma (\theta)^4[1-\beta(\theta)\cos(\alpha_\mathrm{obs}) ]^3 },
\label{con:egamma}
\end{equation}
where $\epsilon(\theta)$ is the normalized angular energy distribution satisfying $\int \epsilon(\theta)\mathrm{d}\Omega = 1$, $\Gamma(\theta)$ is the Lorentz factor profile, and $\beta(\theta)$ is the corresponding dimensionless velocity. The angle $\alpha_{\mathrm{obs}}$ between the jet element at $(\theta,\phi)$ and the line of sight is given by
\begin{equation}
\cos\alpha_{\mathrm{obs}}=\cos\theta_{\mathrm{obs}}\cos\theta+\sin\theta_{\mathrm{obs}}\sin\theta\cos\phi.
\label{con:cosalpha}
\end{equation}

\begin{figure}
\centering
\includegraphics[width=0.5\columnwidth]{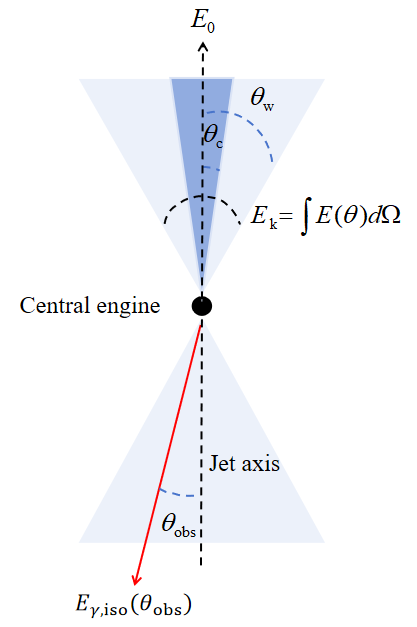}
\caption{Schematic diagram of an off-axis structured jet. The kinetic energy per unit solid angle $E(\theta)$ decreases with polar angle $\theta$ from the jet axis, with an on-axis value $E_0$, a characteristic core angle $\theta_\mathrm{c}$, and a truncation angle $\theta_\mathrm{w}$. Integrating over solid angle yields the total beaming-corrected kinetic energy $E_\mathrm{k}$. An observer located at a viewing angle $\theta_\mathrm{obs}$ measures an isotropic-equivalent gamma-ray energy $E_{\gamma,\mathrm{iso}}$, which depends on relativistic beaming and geometric projection effects.
}
\label{fig:observe}
\end{figure}

Using the jet geometry inferred from afterglow modeling, we normalize Equation~(\ref{con:etheta}) to obtain
\begin{equation}
\epsilon (\theta)=\begin{cases} 
\frac{e^{-\theta^2/2\theta_{\mathrm{c}}^2}}{2\pi\theta_{\mathrm{c}}^2\left[1-
e^{-\theta_{\mathrm{w}}^2/2\theta_{\mathrm{c}}^2}\right]},   & \mathrm{for \ \ }\theta  \leq \theta_{\mathrm{w}}, \\
0, &  \mathrm{for \ \ }\theta  > \theta_{\mathrm{w}},
\end{cases}
\label{con:epsilon1}
\end{equation}
and assume that the Lorentz factor follows the same Gaussian structure as
\begin{equation}
\Gamma(\theta)=\begin{cases} 
(\Gamma_{\mathrm{0}}-1)e^{-\theta^2/2\theta_{\mathrm{c}}^2}+1, & \mathrm{for \ \ }\theta  \leq  \theta_{\mathrm{w}}, \\
1, &  \mathrm{for \ \ }\theta  > \theta_{\mathrm{w}}.
\end{cases}
\label{con:lorentz}
\end{equation}
For an off-axis Gaussian jet, integrating Equation~(\ref{con:egamma}) over the azimuthal angle yields (\citealt{msr+22})
\begin{equation}
E_{\gamma,\mathrm{iso}}(\theta_{\mathrm{obs}})=\frac{1}{2}  \pi E_{\gamma}\int_{0}^{\theta_{\mathrm{w}}}
\mathrm{d}\theta \frac{(2a^2+b^2)\epsilon (\theta)\sin\theta}{\Gamma (\theta)^4(a^2-b^2)^{5/2}},
\label{con:egammatoiso}
\end{equation}
where $a=1-\beta(\theta)\cos\theta_{\mathrm{obs}}\cos\theta$ and $b=\beta(\theta)\sin\theta_{\mathrm{obs}}\sin\theta$. The factor of $1/2$ accounts for the two symmetric jets.
Note that configurations with $\theta_{\rm obs}>\theta_{\rm w}$ are physically allowed, which correspond to off-axis observers located outside the truncated energetic region of the jet. Although no emitting material lies directly along the line of sight, radiation from $\theta\leq\theta_{\rm w}$ becomes visible as the jet decelerates and its relativistic beaming cone widens. 
A schematic diagram illustrating the jet geometry and key parameters is shown in Figure~\ref{fig:observe}.

In summary, we adopt a self-consistent framework that connects the prompt emission and afterglow properties of sGRBs within the context of a structured jet model. The total jet energy is decomposed into kinetic and radiative components, with the kinetic energy constrained through afterglow modeling and the radiative energy inferred from the prompt emission. By adopting a Gaussian jet structure and accounting for relativistic beaming and viewing angle effects, we relate the observed isotropic-equivalent $\gamma$-ray energy to the intrinsic jet energetics. This framework enables us to consistently derive the key physical parameters required to estimate the $\gamma$-ray efficiency and, ultimately, the central engine properties. In the following subsection, we describe in detail how the afterglow light curves are modeled to constrain the jet geometry and kinetic energy.

\subsection{Afterglow fitting}
\label{sub:afterglowpy}

We use the \texttt{afterglowpy} package to calculate the emission from the FS \citep{rvp+20}. \texttt{Afterglowpy} is a publicly available Python package designed to model GRB afterglows within the structured jet synchrotron emission framework. It computes broadband afterglow light curves and spectra arising from synchrotron radiation emitted by relativistic electrons accelerated in the external FS as the GRB jet decelerates in an interstellar-medium (ISM)–like environment. The code solves the radiative transfer problem by integrating the specific intensity over equal-arrival-time surfaces, fully accounting for relativistic beaming, light-travel-time effects, and jet geometry.

In addition to the classical top-hat model, \texttt{afterglowpy} can handle structured jets such as Gaussian and power-law profiles that have been proposed in previous studies \citep{ajm+05,dm+13,max+18,ccr+20}. When dealing with structured jets, it adopts the single-shell approximation and divides the jet into several annuli characterized by their polar angles ($\theta$). Assuming each annulus evolves independently, \texttt{afterglowpy} employs a semi-analytic formalism to calculate the dynamical evolution of each annulus and then integrates over all contributions to obtain the total observed flux.
However, \texttt{afterglowpy} initializes the blast-wave dynamics from the Blandford--McKee self-similar solution \citep{1976PhFl...19.1130B} and does not explicitly include the preceding coasting phase with an approximately constant Lorentz factor. We assess the impact of this limitation on our results in Appendix~\ref{appendix:vegas}.

For the Gaussian jet model, the set of free parameters used in our Markov chain Monte Carlo (MCMC) fitting is $\boldsymbol{\Theta} = [\theta_{\mathrm{w}}, \theta_{\mathrm{obs}}, E_{0}, \theta_{\mathrm{c}}, n_{0}, p, \epsilon_{\mathrm{e}}, \epsilon_{\mathrm{b}}]$, where $\theta_{\mathrm{w}}$ denotes the truncation angle beyond which the energy drops to zero, $\theta_{\mathrm{obs}}$ is the viewing angle, $E_{\mathrm{0}}$ is the on-axis kinetic energy, $\theta_{\mathrm{c}}$ represents the characteristic half-opening angle of the jet core, $n_{0}$ is the number density of the uniform ISM, $p$ is the power-law index of the electron energy distribution, and $\epsilon_{\mathrm{e}}$ and $\epsilon_{\mathrm{b}}$ represent the fractions of shock energy transferred to electrons and magnetic fields, respectively. In addition, since the fraction of electrons accelerated at the shock front ($\xi_{\mathrm{N}}$) is degenerate with $E_{0}$ and $\epsilon_{\mathrm{e}}$ \citep{ew+05,lhw+22}, we fix $\xi_{\mathrm{N}} = 1$ for simplicity. 
As $\xi_{\mathrm{N}}$ is degenerate with the other afterglow parameters, the same afterglow emission can be reproduced by rescaling the parameters as $E_{0}\rightarrow E_{0}/\xi_{\mathrm{N}}$, $n_{0}\rightarrow n_{0}/\xi_{\mathrm{N}}$, $\epsilon_{\mathrm{e}} \rightarrow\xi_{\mathrm{N}} \epsilon_{\mathrm{e}}$, and $\epsilon_{\mathrm{b}} \rightarrow \xi_{\mathrm{N}} \epsilon_{\mathrm{b}}$. Therefore, adopting a smaller $\xi_{\mathrm{N}}$ would increase the inferred kinetic energy approximately as $E_{\mathrm{k}}\propto\xi_{\mathrm{N}}^{-1}$. Since $\xi_{\mathrm{N}}\leq1$, our choice of $\xi_{\mathrm{N}}=1$ gives the lowest kinetic energy allowed by this degeneracy, and the $E_{\mathrm{k}}$ values reported in this work can therefore be regarded as lower limits in this respect. 

To obtain the posterior distributions of the model parameters $\boldsymbol{\Theta}$, we adopt a Bayesian inference framework
\begin{equation}
p(\boldsymbol{\Theta}|\mathrm{Data}) \propto p(\mathrm{Data}|\boldsymbol{\Theta})\, p(\boldsymbol{\Theta}),
\label{con:bayesian}
\end{equation}
where $p(\mathrm{Data}|\boldsymbol{\Theta})$ is the likelihood function and $p(\boldsymbol{\Theta})$ denotes the prior probability distribution of the model parameters. The likelihood function is constructed assuming Gaussian-distributed measurement uncertainties and is expressed in terms of the observed flux density $F_{\nu,i}$, its associated uncertainty $\sigma_i$, and the model-predicted flux density $F_{\nu,i}(\boldsymbol{\Theta})$ as
\begin{equation}
\ln p(\mathrm{Data}|\boldsymbol{\Theta}) =
-\frac{1}{2} \sum_i
\left[
\frac{F_{\nu,i} - F_{\nu,i}(\boldsymbol{\Theta})}{\sigma_i}
\right]^2 .
\label{con:likelihood}
\end{equation}
In addition, we incorporate the $3\sigma$ upper limits by assigning zero probability to fluxes that exceed these limits. The adopted priors for the \texttt{afterglowpy} model parameters, together with their functional forms and parameter bounds, are summarized in Table~\ref{tab:prior}. 
We note that the upper bound $\log\epsilon_{\mathrm{e}}=-0.5$ is adopted as equipartition-motivated maxima \citep{2016MNRAS.462.1111D}, and should be regarded as physically motivated but informative priors rather than strict limits.
However, extending this bound to $\log\epsilon_{\mathrm{e}}=0$ produces only a modest change in the inferred energetics and does not affect our main conclusions.
Posterior sampling is performed using the affine-invariant MCMC sampler \texttt{emcee} with the Geweke statistic as a convergence diagnostic \citep{g+92,fh+13}. 

\begin{table}
\caption{Parameter estimation priors and bounds used in the MCMC parameter estimation for the Gaussian structured jet model.}
\label{tab:prior}
\centering
\begin{tabular}{lccc}
\hline\hline
Parameters & Unit & Prior form & Bounds \\
\hline
$\theta_{\mathrm{w}}$ & rad & uniform & $(\theta_{\mathrm{c}}, \min\left\{5\theta_{\mathrm{c}},\pi/2\right\}]$ \\
$\theta_{\mathrm{obs}}$ & rad & uniform & $(0,0.8]$ \\
$E_0$ & erg & log-uniform & $[48,55]$ \\
$\theta_{\mathrm{c}}$ & rad & uniform & $[0.01,0.8]$ \\
$n_0$ & $\mathrm{cm^{-3}}$ & log-uniform & $[-6,1]$ \\
$p$ & - & uniform & $(2,5]$ \\
$\epsilon_{\mathrm{e}}$ & - & log-uniform & $[-5,-0.5]$ \\
$\epsilon_{\mathrm{b}}$ & - & log-uniform & $[-5,-0.5]$ \\
\hline
\end{tabular}
\end{table}
\subsection{Fitting results}
\label{sub:results}

Since our primary goal is to investigate the energy distribution of GRBs, we first perform an MCMC fit for each GRB and apply two criteria to the posterior distributions of $E_{\mathrm{0}}$ and $\theta_{\mathrm{c}}$, i.e., the parameters most relevant to the beaming-corrected kinetic energy, in order to identify the ``strictly constrained'' (SC) GRBs as follows.
\begin{enumerate}
\item The widths of the 68\% credible interval of $E_{\mathrm{0}}$ and $\theta_{\mathrm{c}}$ are less than half of the prior range.
\item The peak of the posterior distribution lies within the 68\% credible interval.
\end{enumerate}
These requirements help guarantee that the derived energy-related parameters are well constrained by the data rather than by the choice of priors, which is essential for a reliable characterization of the GRB energy distribution. After applying these criteria, we obtain a set of 15 strictly constrained (SC) GRBs\footnote{For GRB 050709, the posterior distribution is bimodal, and we therefore apply the selection criteria to both peaks.}. The posterior distributions of the model parameters $\boldsymbol{\Theta} = [\theta_{\mathrm{w}}, \theta_{\mathrm{obs}}, E_{0}, \theta_{\mathrm{c}}, n_{0}, p, \epsilon_{\mathrm{e}}, \epsilon_{\mathrm{b}}]$ are summarized in Table~\ref{tab:well}. And the broadband afterglow fits for the 15 SC GRBs are presented in Figure \ref{fig:well}.

For the remaining 25 bursts, in order to obtain a well-constrained energy distribution while maintaining a consistent fitting procedure, we adopt an empirical prior approach. Instead of using the initial uniform priors, we construct the prior distributions of the microphysical and jet structure parameters, including $\epsilon_{\mathrm{e}}$, $\epsilon_{\mathrm{b}}$, and the ratio $\theta_{\mathrm{w}}/\theta_{\mathrm{c}}$, from the posterior samples of the 15 SC GRBs. These empirically derived distributions are then used as priors for the poorly constrained GRBs. This approach extends previous studies (e.g., \citealt{fbm+15,wrj+24}), which assumed identical parameter values across all GRBs, by using population-informed priors that allow burst-to-burst variations, thereby providing a more flexible and realistic description of the GRB population. Although this approach did not give tight constraints on $\theta_{\mathrm{c}}$, it did provide tighter constraints on the on-axis energy $E_0$ for all GRBs. The posterior distributions of the remaining GRBs are summarized in Table~\ref{tab:other}. Figure~\ref{fig:others} shows the broadband light curves of the remaining 25 GRB afterglows, following the same plotting conventions as Figure~\ref{fig:well}. We note that a more comprehensive treatment using a hierarchical Bayesian model may provide more realistic constraints \citep{l+13,tt+19,mfg+19}. Nevertheless, owing to the substantial computational cost, we do not perform a full hierarchical analysis. Instead, we approximate the posterior distributions of the afterglow parameters using empirical prior functions, which serve as the basis for the following statistical analysis.

\subsection{Energetics distribution}
\label{sub:energetics}
\begin{figure}
\centering
\includegraphics[width=0.99\columnwidth]{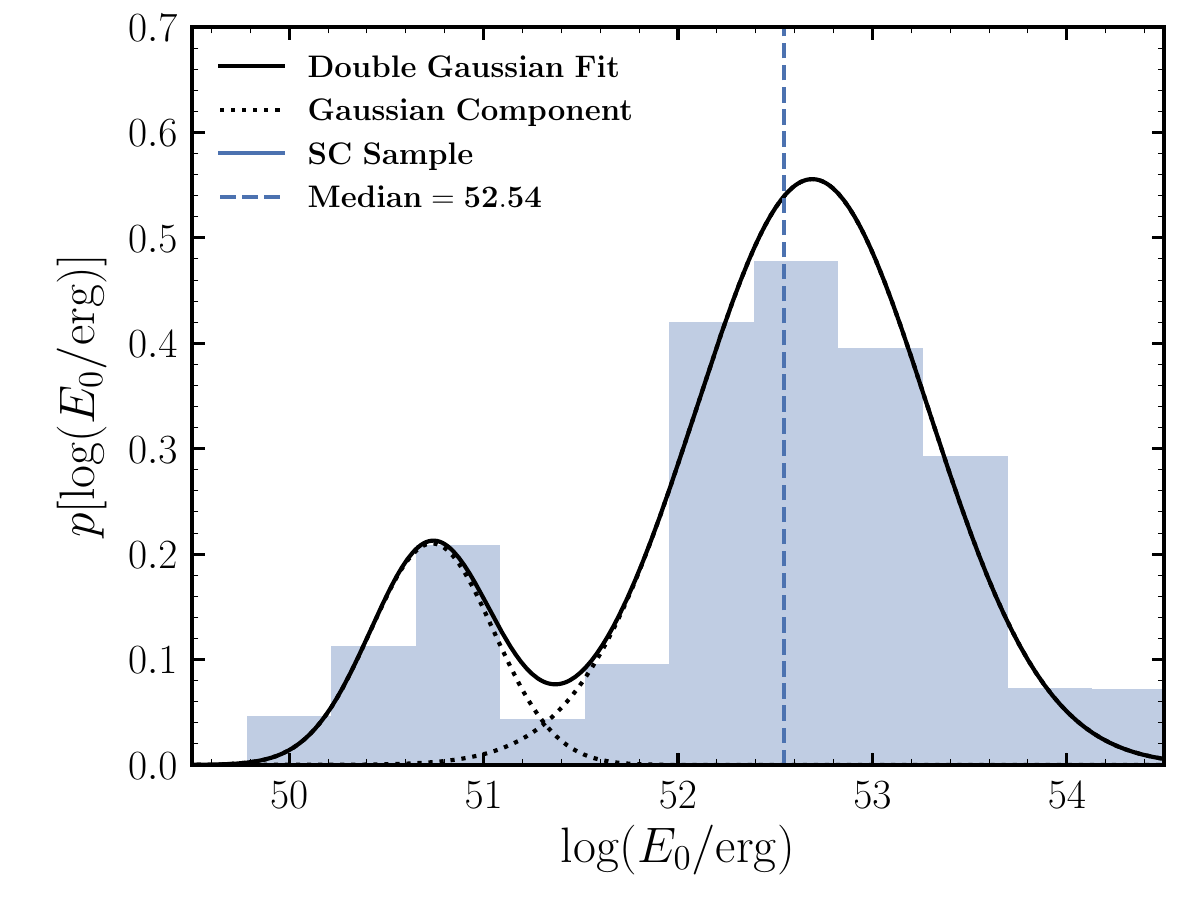}
\caption{
Posterior distributions of $E_0$ for the SC sample, shown in blue shaded histogram, with the median values indicated. 
The black solid line represents the best-fitting double-Gaussian model for the SC sample, as described in Equation~(\ref{con:E0}), while the black dotted lines denote the individual Gaussian components.
}
\label{fig:E0}
\end{figure}

Since the model parameters of the SC bursts are better constrained, we focus on the statistical results derived from the SC sample in the main text, while the corresponding results for the full sample are presented in Appendix~\ref{appendix:full}.
For each GRB, we randomly select 6000 samples from the posterior distribution of $E_0$ and construct a normalized probability density distribution using all GRBs. The median value of $\log E_0$ is 52.54, with a 68\% credible interval of [51.08, 53.43]. We then fit the resulting distribution using a double-Gaussian model, whose functional form is given in
\begin{equation}
\begin{split}
p(\log E_0)=A_1\frac{1}{\sqrt{2\pi}\sigma_1}\exp\left\{{-\frac{[\log E_0-\mu_1]^2}{2\sigma_1^2}}\right\}+
\\(1-A_1)\frac{1}{\sqrt{2\pi}\sigma_2}\exp\left\{{-\frac{[\log E_0-\mu_2]^2}{2\sigma_2^2}}\right\}. 
\end{split}
\label{con:E0}
\end{equation} 
The best-fit values of the model parameters are $A_1=0.17$, $\mu_1=50.74$, $\sigma_1=0.44$, $\mu_2=52.70$, and $\sigma_2=0.60$. Figure~\ref{fig:E0} shows the distributions of the SC sample. The best-fits of the double-Gaussian model for the SC sample are also presented.
The double-Gaussian model broadly reproduces the overall shape of the inferred distribution, including the dominant peak and the lower-energy enhancement. Although some deviations remain, the model provides a reasonable approximation to the distribution.

\begin{figure}
\centering
\includegraphics[width=0.99\columnwidth]{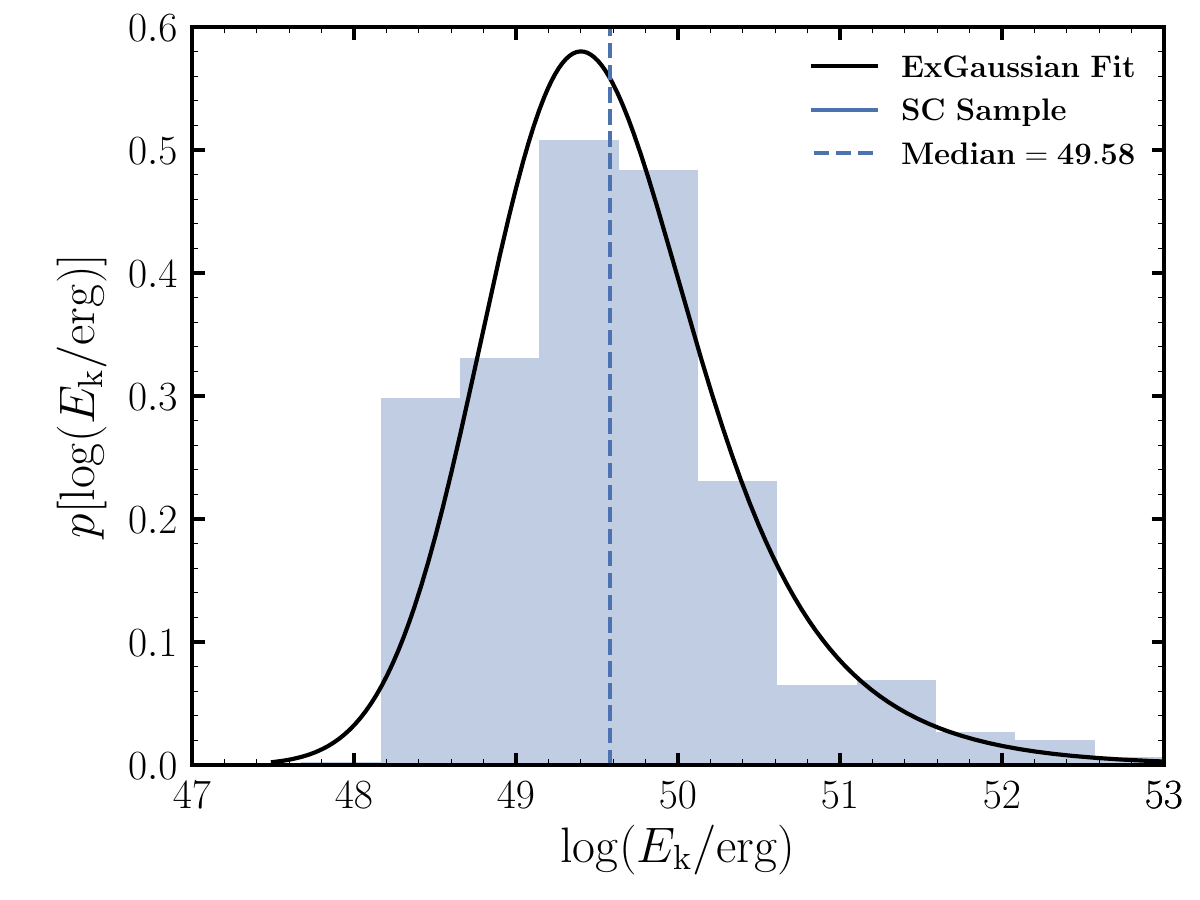}
\caption{
Posterior distributions of $E_\mathrm{k}$ for the SC sample, shown in blue shaded histogram, with the median values indicated. 
The black solid line represents the best-fitting ExGaussian model for the SC sample, as described in Equation~(\ref{con:ex}).
}
\label{fig:Ek}
\end{figure}

We then calculate the total kinetic energy $E_\mathrm{k}$ following the same procedure as for $E_0$. For each GRB, we randomly draw 6000 samples from the joint posterior distribution of [$E_0$, $\theta_\mathrm{c}$, $\theta_\mathrm{w}$]. Each set of sampled parameters is then converted to a corresponding value of $E_\mathrm{k}$ through Equation~(\ref{con:ek}), allowing us to construct the posterior distribution of log($E_\mathrm{k}$) for the SC sample. The median value of $\mathrm{log}(E_\mathrm{k})$ is 49.58, with a 68\% credible interval of [48.71, 50.28].
We model the resulting distribution using an exponentially modified Gaussian (ExGaussian) model, which allows for an asymmetric distribution with an extended high-energy tail. The functional form is given by
\begin{equation}
\begin{aligned}
p(\log E_\mathrm{k}) = \frac{1}{2\tau}
\exp\left[
\frac{\sigma^2}{2\tau^2}
-\frac{\log E_\mathrm{k}-\mu}{\tau}
\right]
\operatorname{erfc}\left(
\frac{\mu+\sigma^2/\tau-\log E_\mathrm{k}}{\sqrt{2}\sigma}
\right),
\end{aligned}
\label{con:ex}
\end{equation}
where $\mu$ and $\sigma$ are the location and standard deviation of the Gaussian component, respectively, and $\tau$ is the scale parameter of the exponential component. The best-fit parameters are $\mu=49.07$, $\sigma=0.51$, and $\tau=0.59$. Figure~\ref{fig:Ek} shows the combined posterior distribution, the best-fitting ExGaussian model, and the corresponding histogram.

In this work, the total kinetic energy of each GRB has been obtained by the afterglow modeling, which enables an estimate of the radiative efficiency according to Equations~\eqref{con:eta} and \eqref{con:egamma}. However, an estimate of the total energy emitted in $\gamma$ rays requires knowledge of the initial Lorentz factor $\Gamma_0$. The value of $\Gamma_0$ generally depends on detailed modeling of the whole light curve including the plateau phase and can range from tens to hundreds, depending on the adopted model \citep{ lyz+10,sm+12,dhp+22,dhk+25}. In the absence of a direct constraint for individual bursts, we adopt a fiducial value of $\Gamma_0 = 300$. For different $\Gamma_0$, we discuss later in Section~\ref{sec:discussion}. 

\begin{figure}
\centering
\includegraphics[width=0.99\columnwidth]{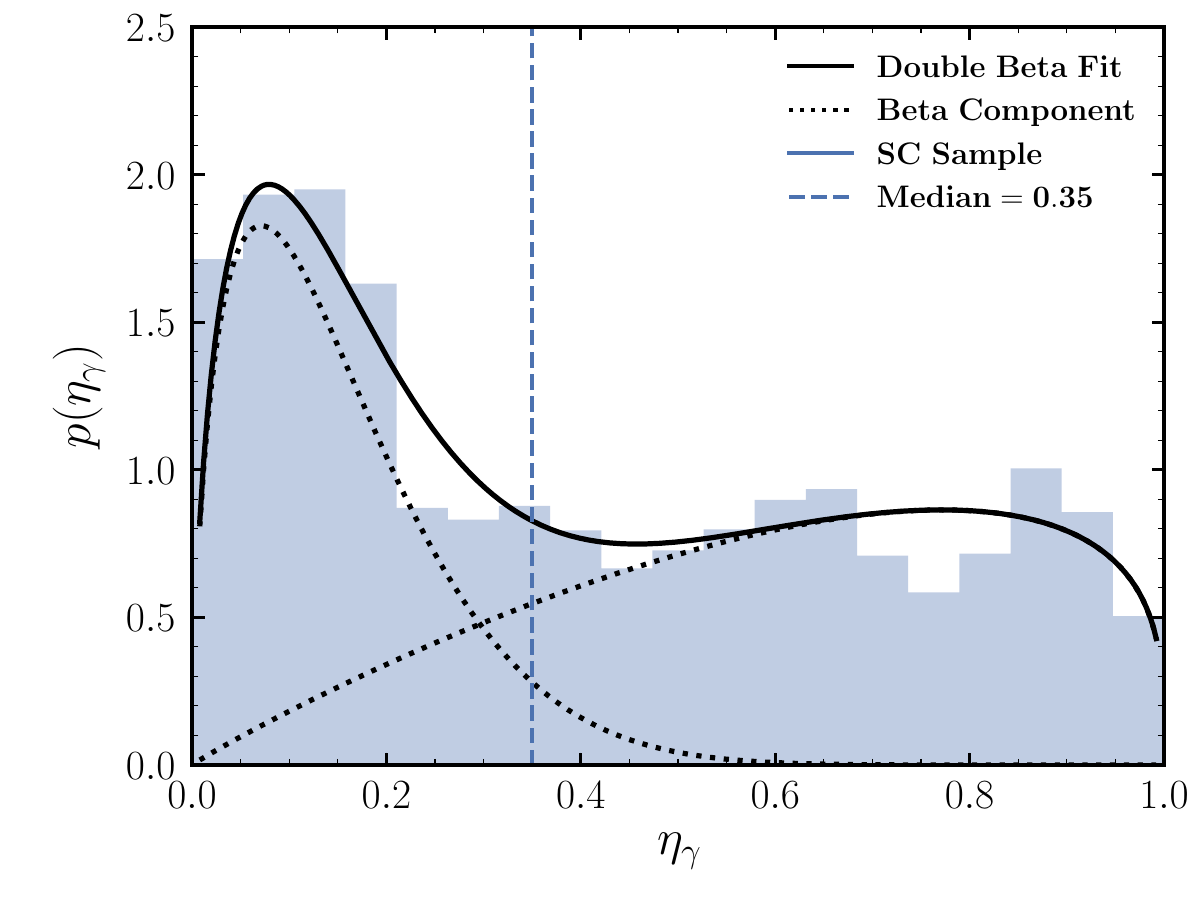}
\caption{
Posterior distributions of $\eta_\gamma$ for the SC sample, shown in blue shaded histogram, with the median values indicated. The black solid line represents the best-fitting double-Beta model for the SC sample, as described in Equation~(\ref{con:beta}), while the black dotted lines denote the individual Beta components.}
\label{fig:eta}
\end{figure}

After fixing $\Gamma_0$, the prompt emission energy $E_\gamma$ is estimated following the same procedure adopted for the kinetic energy $E_{\mathrm{k}}$. The isotropic-equivalent prompt energy $E_{\gamma,\mathrm{iso}}$ is derived by modeling the observed fluence $f_\gamma$ as a Gaussian-distributed quantity using the median value and standard deviation reported in Table~\ref{tab:redshift}. For each burst, we draw samples from the joint posterior distribution of the jet parameters [$E_0$, $\theta_{\mathrm{c}}$, $\theta_{\mathrm{w}}$, $\theta_{\mathrm{obs}}$] and convert them to $E_\gamma$. The resulting distribution has a median value of $\log(E_\gamma)=49.43$, with a 68\% credible interval of [48.15, 50.30]. By combining the posterior distributions of $E_\gamma$ and the corresponding $E_{\mathrm{k}}$ for each burst, we obtain the posterior distribution of the radiative efficiency $\eta_\gamma$. For each posterior sample, $\eta_\gamma$ is computed using Equation~\eqref{con:eta}, thereby preserving the covariance between $E_{\mathrm{k}}$ and $E_\gamma$ through the joint sampling procedure. The resulting efficiency distribution has a median value of $\eta_\gamma = 0.35$, with a 68\% credible interval of [0.09, 0.79]. We further model the distribution of $\eta_\gamma$ using a double-Beta distribution,
\begin{equation}
\begin{split}
p(\eta_\gamma) = A_1\frac{\Gamma(\alpha_1+\beta_1)}{\Gamma(\alpha_1)\Gamma(\beta_1)} 
\, \eta_\gamma^{\alpha_1-1} (1-\eta_\gamma)^{\beta_1-1}+
\\(1-A_1)\frac{\Gamma(\alpha_2+\beta_2)}{\Gamma(\alpha_2)\Gamma(\beta_2)} 
\, \eta_\gamma^{\alpha_2-1} (1-\eta_\gamma)^{\beta_2-1},
\end{split}
\label{con:beta}
\end{equation}
for which the best-fit values for the model parameters are $\alpha_1 = 1.61, \beta_1 = 8.99, \alpha_2 = 1.95, \beta_2 = 1.28$, and $A_1 = 0.41$. Figure~\ref{fig:eta} presents the posterior distribution of the radiative efficiency $\eta_\gamma$, along with its median value. We restrict the fitting range to $\eta_\gamma \in [0.1\%,99.9\%]$ to avoid numerical instabilities of the Beta function near the boundaries at 0 and 1. Over this interval, the relative deviations between the fitted double-Beta distribution and the inferred posterior remain below approximately $30\%$ across the entire range of $\eta_\gamma$, indicating that the double-Beta function provides a satisfactory representation of the radiative efficiency distribution.

Notably, \citet{fbm+15} performed a similar analysis using a top-hat jet model without incorporating viewing-angle effects. In their study, the inferred on-axis kinetic energy $E_0$ is typically $\sim2\times10^{51}$ erg, which is about an order of magnitude lower than the value obtained in our Gaussian-jet modeling ($\sim3.5\times10^{52}$ erg). This discrepancy naturally arises from their assumption that all bursts are viewed exactly on-axis: when the observer is located away from the jet core, a substantially larger intrinsic energy is required to reproduce the same observed afterglow flux, which our model captures through the inclusion of viewing geometry.
After accounting for the beaming correction and the full viewing-angle dependence, the kinetic energy inferred in our study is in good agreement with that reported in theirs ($4\times10^{49}$\,erg versus $8\times10^{49}$\,erg). The inferred radiative efficiencies are also comparable, with median values of $\eta_\gamma \approx 0.35$ in our study and $\eta_\gamma \approx 0.40$ in theirs.

\section{On the magnetic field of BNS merger BH}
\label{sec:BZ}

The jet energetics derived above provide valuable information about the physical conditions of the central engine. In the case of BNS mergers, the remnant is expected to consist of a rapidly rotating BH surrounded by a dense accretion disk, whose magnetic field plays a crucial role in the launching and evolution of relativistic jets. In particular, it regulates the level of baryon pollution in the polar region, enables the launching of relativistic outflows with Lorentz factors of order a few hundred, and facilitates the unbinding of matter from the post-merger remnant torus which plausibly contributes the bulk of the NS merger ejecta \citep{mrd+19,n+20,pac+22}. Despite its importance, the strength of the magnetic field associated with the post-merger BH remains poorly constrained. On the one hand, direct observational measurements of BH magnetic fields are currently unavailable. On the other hand, numerical simulations of BNS mergers have produced a wide range of predictions for the maximum magnetic field strength attainable during and after the merger. These discrepancies arise primarily from the unresolved amplification of magnetic fields by magnetohydrodynamic instabilities such as the Kelvin–Helmholtz instability (KHI) during the merger phase \citep{zm+13,kck+15,gzd+15,avp+22} and the magnetorotational instability (MRI) in the post-merger accretion disk \citep{bh+91,bh+98,ksk+15,kks+18}, as well as from a strong sensitivity to the assumed initial condition \citep{gbj+17}. As a result, the magnetic field properties of post-merger BHs remain subject to substantial theoretical and numerical uncertainties.

Although several central engine mechanisms have been proposed for sGRBs, including spin-down-powered outflows from a rapidly rotating millisecond magnetar and jets powered by neutrino--antineutrino annihilation above a hyperaccreting disk \citep{mqt+08,zb+11}, the BZ mechanism provides a particularly direct connection between the jet energetics and the properties of the central BH. In this framework, a rotating Kerr BH threaded by large-scale magnetic fields can extract rotational energy electromagnetically and power a relativistic jet \citep{bz+77}. The BZ process has therefore been widely considered as a promising mechanism for producing sGRBs. In particular, \citet{lfm+19} explored the connection between GRB energetics and the magnetic field threading the BH horizon within the BZ framework and inferred characteristic magnetic field strengths in the range of $10^{15}-10^{17}~\mathrm{G}$. While their work represents an important step toward constraining the magnetic field of GRB central engines, their analysis relied on simplified assumptions regarding the GRB energy distribution and adopted constant BH mass and spin parameters.

Motivated by these considerations, we estimate the magnetic field strength of BNS merger remnant BHs using the jet energy distribution derived above, together with realistic distributions of BH mass and spin. We adopt the BZ mechanism as a working hypothesis because it provides a direct quantitative mapping between the observed jet energetics and the magnetic field at the BH horizon. The magnetic field constraints derived below are therefore conditional on a predominantly BZ-powered jet and do not exclude alternative central engine scenarios.
In Section~\ref{sub:BZ}, we briefly introduce the BZ process. In Section~\ref{sub:BHmass}, we describe the method used to estimate the BH mass and spin and to infer the corresponding magnetic field distributions. The inferred magnetic field properties and their physical implications are then presented and discussed in Section~\ref{sub:BZresults}.

\subsection{Blandford–Znajek process}
\label{sub:BZ}

The BZ process provides a mechanism for extracting rotational energy from a spinning BH to power relativistic jets. In this mechanism, magnetic field lines anchored in the surrounding plasma thread the event horizon of the BH and are forced to rotate due to frame dragging. The resulting electromagnetic torque transfers angular momentum and energy from the BH to the outflow, producing a Poynting-flux-dominated jet \citep{mt+82}. Since its proposal, the BZ process has been extensively studied using both analytical models and numerical simulations. Previous works have investigated the energetics, collimation, and morphology of BZ-powered jets, as well as the efficiency of the process and its dependence on BH spin and accretion properties \citep{lbw+00,wym+04,mj+05,mj+06,tnm+11,hlm+24,wdb+25}. In addition, the coupled evolution of BH mass and spin under the action of electromagnetic torques associated with BZ jets has also been explored \citep{mtb+12,ll+24}.

In the context of BZ process, the jet luminosity can be related to the properties of the central BH \citep{tnm+11, tg+15} as:
\begin{equation}
L_\mathrm{BZ}
=\frac{\kappa}{4\pi~c}\Omega_\mathrm{H}^2\Phi_\mathrm{BH}^2f(\Omega_\mathrm{H}),
\label{con:LBZ}
\end{equation}
where $\kappa$ is a numerical constant depending on the magnetic field geometry (of order $\sim0.05$), $\Omega_\mathrm{H} =ac/2r_{\mathrm{H}}$ is the angular frequency of the BH horizon, and $\Phi_\mathrm{BH}$ denotes the magnetic flux threading the horizon of one hemisphere. Here $a$ is the dimensionless BH spin parameter, $r_{\mathrm{H}}=r_{\mathrm{g}}(1+\sqrt{1-a^2})$ is the radius of the horizon, and $r_\mathrm{g}=GM_\mathrm{BH}/c^2$ is the gravitational radius of the BH. The factor $f(\Omega_\mathrm{H})$ accounts for higher-order corrections at moderate to high spin and can be approximated by a sixth-order expansion \citep{tan+10} as $f(\Omega_\mathrm{H})\approx1+1.38(\Omega_\mathrm{H}r_\mathrm{g}/c)^2- 9.2(\Omega_\mathrm{H}r_\mathrm{g}/c)^4$.

For a BZ-powered jet, we assume that the energy extracted via the BZ mechanism is ultimately channeled into the jet, contributing to both the kinetic energy of the outflow and the prompt radiative emission. Under this assumption, the BZ energy can be written as $E_\mathrm{BZ}=E_\mathrm{k}+E_\gamma$, where $E_\mathrm{k}$ and $E_\gamma$ denote the kinetic and radiative energies of the jet, respectively. Assuming a predominantly poloidal magnetic field configuration near the BH horizon, which maximizes the efficiency of jet launching, Equation~(\ref{con:LBZ}) can be recast in terms of the magnetic field strength \citep{lfm+19} as
\begin{equation}
\frac{L_\mathrm{k}}{(1-\eta_\gamma)}
\approx10^{51}\,\mathrm{erg\,s^{-1}}
\left(\frac{a}{0.9}\right)^{2}
\left(\frac{M_\mathrm{BH}}{5\,M_\odot}\right)^{2}
\left(\frac{B_\mathrm{BH}}{10^{16}\,\mathrm{G}}\right)^2 ,
\label{con:B}
\end{equation}
where $\eta_\gamma$ is the radiative efficiency, $a$ is the dimensionless BH spin parameter, $M_\mathrm{BH}$ is the BH mass, $M_\odot$ denotes the solar mass, $B_\mathrm{BH}$ is the magnetic field strength (poloidal) associated with the post-merger accretion disk–BH system at the BH horizon. Here, $L_\mathrm{k}$ is defined as an effective kinetic luminosity, $L_\mathrm{k}\equiv E_\mathrm{k}/T_{90}$, where $T_{90}$ is the time interval over which 90\% of the prompt $\gamma$-ray fluence is detected. We note that the inferred magnetic field strength corresponds to the characteristic field threading the BH horizon and depend on the assumed geometry, thus it does not necessarily reflect the large-scale magnetic field structure of the jet. Equation~(\ref{con:B}) therefore allows us to derive the magnetic field distribution $B_\mathrm{BH}$ from the distributions of $L_\mathrm{k}$, $\eta_\gamma$, $a$, and $M_\mathrm{BH}$. 
The distributions of $\eta_\gamma$ and $E_{\mathrm{k}}$ have been obtained above. For each burst, we adopt its individually measured $T_{90}$ from literature \citep{sll+18,fnd+22,ll+24}, convert the observed duration to the source rest frame as $T_{90,\mathrm{rest}}=T_{90,\mathrm{obs}}/(1+z)$ and calculate the corresponding kinetic luminosity.

\subsection{The estimation of BH mass and spin}
\label{sub:BHmass}

Under the assumption that sGRBs originate from BNS mergers, we construct a synthetic population of merging BNS systems based on the cosmic star-formation history of \citet{md+14} and the metallicity evolution model of \citet{bhb+16}. The BNS population is generated using a binary evolution model that reproduces both the BNS merger rate reported from \cite{2026arXiv260527226T} and the observed orbital-period and eccentricity distributions of Galactic BNS systems detected as radio pulsars. More specifically, the masses of NSs are determined self-consistently within the binary evolution calculations \citep{hpt+00,htp+02,yj+10}. For each progenitor star that undergoes core collapse, the remnant mass is calculated using a remnant-mass prescription that relates the compact-object mass to both the total stellar mass and the carbon--oxygen core mass immediately prior to the supernova explosion \citep{fbw+12}. Combined with the subsequent binary evolution, including mass transfer, common-envelope evolution \citep{w+84,mpg+21,rd+23}, supernova kicks \citep{hll+05,vn+18}, and gravitational-wave-driven orbital decay \citep{p+64}, this procedure yields a self-consistent distribution of NS masses at the time of merger \citep{cyl+22,clm+25}. These merger-time masses are subsequently used as inputs to fitting formulae calibrated from NR simulations to determine the mass and spin of the post-merger BH remnant. For the NS magnetic fields, at birth, each NS is assigned a surface dipole magnetic field according to a progenitor-mass-dependent distribution. Systems originating from lower-mass progenitors are assigned magnetic fields drawn from a log-normal distribution representative of ordinary radio pulsars \citep{fk+06,ip+13}, whereas NSs formed from more massive progenitors are allowed to populate the stronger-field regime associated with magnetars \citep{ok+14,eri+21}. The subsequent evolution of the magnetic field is described by an exponential decay model, with accelerated decay during episodes of mass accretion \citep{csh+21, 2023MNRAS.526.2210S, 2024ApJ...963...80D}. As a result, the magnetic field of each NS at the time of merger reflects both its natal field strength and its binary evolutionary history. These magnetic fields are retained for comparison with the magnetic fields inferred for the remnant BHs, enabling an estimate of the magnetic field amplification during the merger process. Further details of the population-synthesis methodology can be found in \citet{cyl+22,cly+25} and \cite{czl+25,cll+26}.

In BNS mergers, the tidal deformability characterizes the response of NSs to the tidal field of their companion and encodes information about the NS internal structure. The dimensionless tidal deformability $\Lambda$ is defined through the induced quadrupole moment under an external tidal field, and the mass-weighted combination $\tilde{\Lambda}$ naturally enters the GW phase evolution during the inspiral. The range of the dimensionless tidal deformability $\tilde{\Lambda}$ can be expressed as a function of the chirp mass, in units of the solar mass $M_\odot$,
\begin{equation}
M_\mathrm{chirp} = (M_{\mathrm{NS,1}} M_{\mathrm{NS,2}})^{3/5}
(M_{\mathrm{NS,1}} + M_{\mathrm{NS,2}})^{-1/5},
\end{equation}
following \citet{aer+22}:
\begin{equation}
\tilde{\Lambda  } =
\begin{cases}
-50+500M_\mathrm{chirp}^{-4.5} & \tilde{\Lambda  }=\tilde{\Lambda  }_{\mathrm{min}}\\
-20+1800M_\mathrm{chirp}^{-5} & \tilde{\Lambda  }=\tilde{\Lambda  }_{\mathrm{max}}.
\end{cases}
\label{con:lumbda}
\end{equation}
Given $\tilde{\Lambda}$ and the component masses, the mass of the post-merger BH can be estimated using the NR fitting formula of \citet{cdm+19}:
\begin{equation}
M_\mathrm{BH}=0.98\left(\frac{\nu}{0.25}\right)^2\left(M-0.093\frac{\tilde{\Lambda  }}{400}\right), 
\label{con:MBH}
\end{equation}
where $\nu = \frac{M_{\mathrm{NS,1}} M_{\mathrm{NS,2}}}{(M_{\mathrm{NS,1}} + M_{\mathrm{NS,2}})^2}$ is the symmetric mass ratio and $M = M_{\mathrm{NS,1}} + M_{\mathrm{NS,2}}$ is the total binary mass. The dimensionless spin of the remnant BH is then estimated as \citep{cdm+19}
\begin{equation}
a=\mathrm{tanh}[0.537\nu^2(M-0.185\tilde{\Lambda  })-0.514].
\label{con:spin}
\end{equation}
These estimates of the BH mass and spin provide the necessary inputs for evaluating the magnetic field strength associated with the BZ process.

\begin{figure*}
\centering
\includegraphics[width=\textwidth]{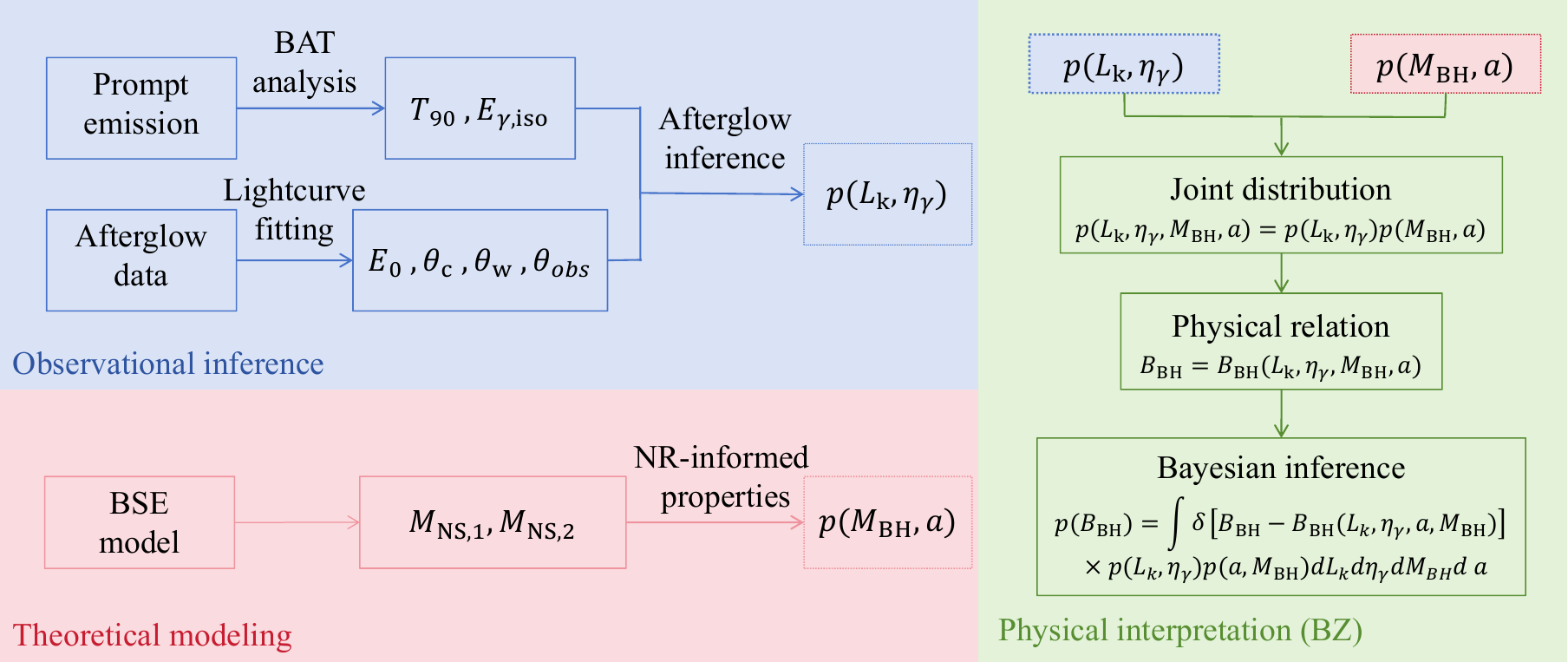}
\caption{Flowchart illustrating the procedure used to infer the magnetic field of the post-merger BH from observational and model inputs.
}
\label{fig:flow}
\end{figure*}

After determining the distributions of the relevant jet and BH parameters, the probability distribution of the BH magnetic field strength $B_\mathrm{BH}$ can be obtained through
\begin{equation}
\begin{split}
p(B_\mathrm{BH})
=&
\int
\delta\!\left[
B_\mathrm{BH}
-
B_\mathrm{BH}(L_\mathrm{k},\eta_\gamma,a,M_\mathrm{BH})
\right]
\\
&\times
p(L_\mathrm{k},\eta_\gamma,a,M_\mathrm{BH})
\,
\mathrm dL_\mathrm{k}
\mathrm d\eta_\gamma
\mathrm dM_\mathrm{BH}
\mathrm da ,
\end{split}
\label{con:B_distribution}
\end{equation}
where the theoretical BH magnetic field strength $B_\mathrm{BH}(L_\mathrm{k},\eta_\gamma,a,M_\mathrm{BH})$ is calculated from Equation~(\ref{con:B}). In principle, evaluating this distribution requires knowledge of the full joint probability distribution $p(L_\mathrm{k},\eta_\gamma,a,M_\mathrm{BH})$, which depends on the detailed relation between the properties of the remnant BH and those of the observed GRB jets. At present, however, neither theoretical models nor observations provide robust constraints on such correlations. Therefore, as a first-order approximation, we assume that the BH properties $(M_\mathrm{BH},a)$ inferred from the synthetic BNS population are statistically independent of the jet properties $(L_\mathrm{k},\eta_\gamma)$ derived from the observational sample. In addition, the distributions of the remnant BH mass and spin produced by BNS mergers span a relatively narrow range compared with the several order-of-magnitude spread of the inferred jet energetics, we expect the uncertainty introduced by this approximation to be sub-dominant compared with that associated with the jet energetics. Under this assumption, the joint distribution can be factorized as
\begin{equation}
p(L_\mathrm{k},\eta_\gamma,a,M_\mathrm{BH})
=
p(L_\mathrm{k},\eta_\gamma)\,
p(a,M_\mathrm{BH}).
\end{equation}
Finally, the BH magnetic field distribution is obtained through Monte Carlo sampling of the jet property and BH property distributions. The overall statistical framework connecting the observational sGRB sample to the synthetic BNS population is illustrated in Figure~\ref{fig:flow}.

We note that our analysis relies on two key assumptions. First, we assume that all GRBs in our sample originate from BNS mergers.
Second, we assume that the synthetic BNS systems form BHs surrounded by sufficiently massive disks to power relativistic jets. We initially require $M_{\mathrm{BH}}>2.5M_\odot$ to exclude stable NS remnants (this treatment is consistent with \citealt{csh+21}). To reduce possible contamination from prompt-collapse systems with negligible disks, we additionally impose a conservative upper limit of $M_{\mathrm{BH}}<2.8M_\odot$. This upper limit is used as a practical proxy rather than a universal prompt-collapse threshold, which depends on the equations of state (EOSs) and binary properties \citep{2013PhRvL.111m1101B,2016MNRAS.459..646B}.
The effects of these assumptions, together with the sensitivity of the inferred magnetic field distribution to the adopted BH mass thresholds and NS EOSs, are discussed in Section~\ref{sec:discussion}.

\subsection{Results and implications}
\label{sub:BZresults}

Since the model parameters of the SC bursts are better constrained, we focus on the magnetic field results derived from the SC sample in this section, while the corresponding results for the full sample are presented in Appendix~\ref{appendix:full}.
To better illustrate our results, we define a dimensionless parameters, $\eta_B$. This is defined as $\eta_B \equiv B_\mathrm{BH}/B_\mathrm{NS}$, which quantifies the amplification of the magnetic field strength from the NS stage to the post-merger BH. 

\begin{figure}
\centering
\includegraphics[width=0.99\columnwidth]{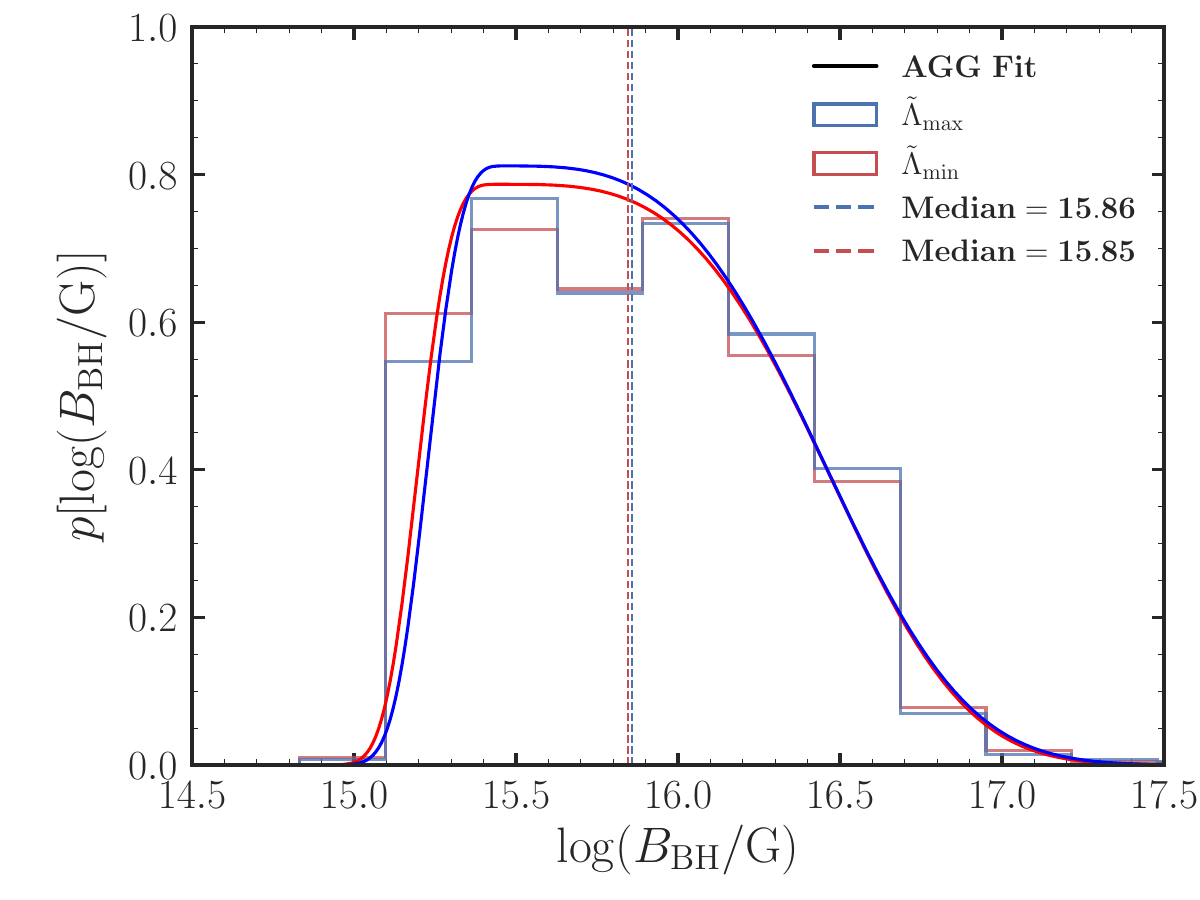}
\includegraphics[width=0.99\columnwidth]{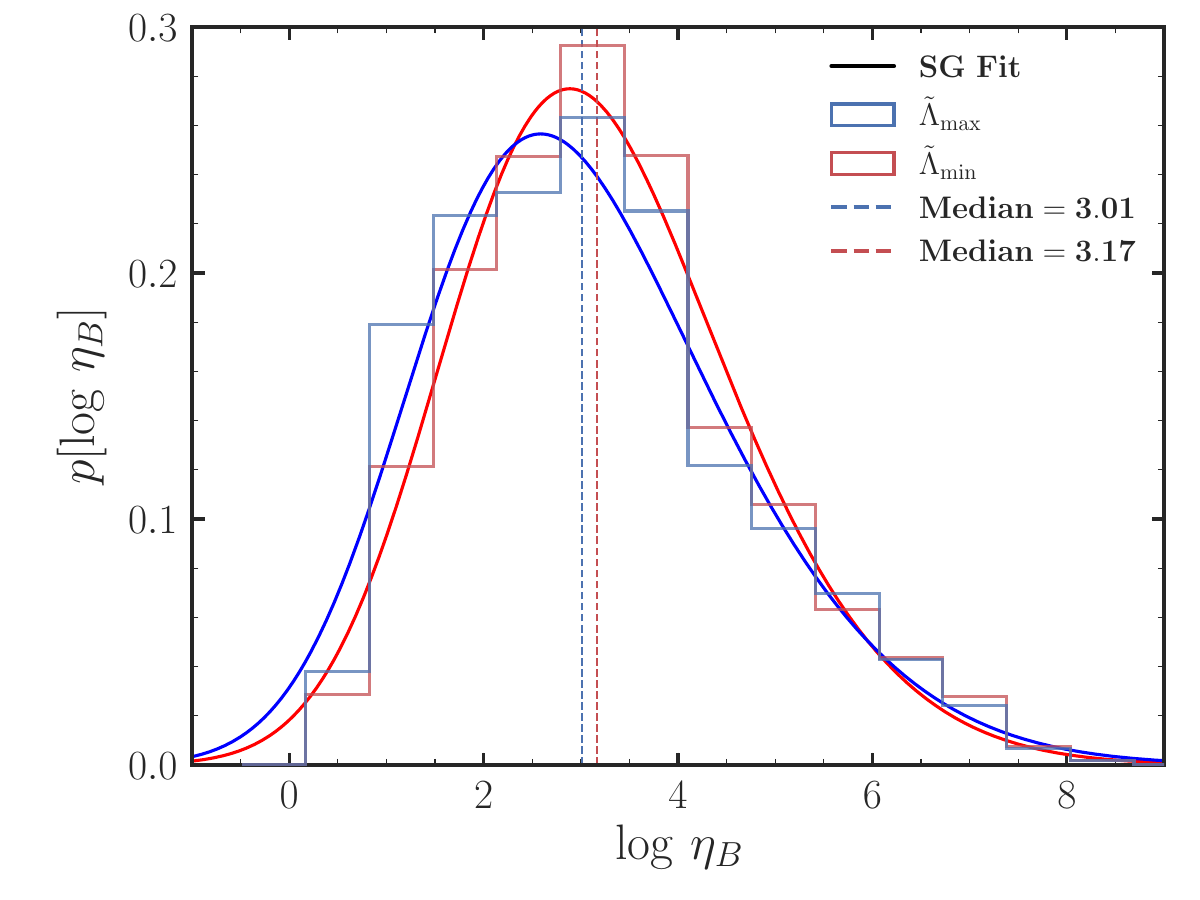}
\caption{Inferred magnetic field properties of the central BH within the BZ framework. From top to bottom, the panels display the distributions of the BH magnetic field strength $\log B_\mathrm{BH}$ and the magnetic field amplification factor $\log \eta_B$, respectively. The red and blue histograms correspond to the results obtained by adopting $\tilde{\Lambda}_\mathrm{min}$ and $\tilde{\Lambda}_\mathrm{max}$ in Equation~(\ref{con:lumbda}).
Vertical dashed lines indicate the corresponding median values, respectively. Solid curves show the fits to the distributions: an AGG model for the magnetic field strength (top) and an SG model for the amplification factor (bottom).
}
\label{fig:B}
\end{figure}

Figure~\ref{fig:B} shows the inferred distributions of the BH magnetic field strength $B_\mathrm{BH}$ (top panel) and the logarithmic amplification factor $\log \eta_B$ (bottom panel). Although only the median values are displayed, the 68\% credible intervals are computed from the full posterior distributions. For $\tilde{\Lambda}_{\mathrm{max}}$, the values are $B_\mathrm{BH}$ = $7.2^{+16.3}_{-4.9}\times10^{15}$\,G and $\eta_B$ = $1026^{+53596}_{-986}$. For $\tilde{\Lambda}_{\mathrm{min}}$, we obtain $B_\mathrm{BH}$ = $7.1^{+15.8}_{-4.8}\times10^{15}$\,G and $\eta_B$ = $1463^{+57899}_{-1401}$. 
To characterize the $B_\mathrm{BH}$ distributions under different $\tilde{\Lambda}$ assumptions, we adopt an asymmetric generalized Gaussian (AGG) distribution, which allows for different widths on either side of the peak. The functional form is given by
\begin{equation}
p(\log B_\mathrm{BH})=\frac{\beta}{(\sigma_L+\sigma_R)\Gamma(1/\beta)}
\exp\left[
-\left(\frac{|\log B_\mathrm{BH}-\mu|}{\sigma(\log B_\mathrm{BH})}\right)^\beta
\right],
\label{con:AGG}
\end{equation}
where
\begin{equation}
\sigma(\log B_\mathrm{BH})=
\begin{cases}
\sigma_L, & \log B_\mathrm{BH}<\mu,\\
\sigma_R, & \log B_\mathrm{BH}\geq\mu.
\end{cases}
\end{equation}
Here $\mu$ is the location parameter, $\sigma_L$ and $\sigma_R$ characterize the widths on the left and right sides of the peak, respectively, and $\beta$ controls the shape of the distribution. The best-fitting parameters $(\mu,\sigma_L,\sigma_R,\beta)$ are $(15.44,0.27,1.14,3.50)$ for $\tilde{\Lambda}_{\mathrm{min}}$ and $(15.46,0.26,1.11,3.29)$ for $\tilde{\Lambda}_{\mathrm{max}}$.
For $\log \eta_B$, the distributions exhibit significant asymmetry and are therefore modeled using a skew-Gaussian (SG) distribution,
\begin{equation}
p (\log \eta_B)=
\frac{2}{\sigma}
\phi\left(\frac{\log \eta_B-\mu}{\sigma}\right)
\Phi\left(\alpha\frac{\log \eta_B-\mu}{\sigma}\right),
\label{con:skewed}
\end{equation}
where $\mu$ and $\sigma$ denote the location and scale parameters, and $\alpha$ quantifies the skewness; the distribution reduces to a standard Gaussian when $\alpha=0$. The best-fitting parameters are $(\alpha,\mu,\sigma)=(1.75,1.77,2.08)$ for $\tilde{\Lambda}_{\mathrm{min}}$, and $(2.14,1.35,2.36)$ for $\tilde{\Lambda}_{\mathrm{max}}$. The corresponding fits are also presented in Figure~\ref{fig:B}.

Our results indicate that, under the assumption that the BZ process powers the majority of sGRBs, magnetic field strengths of order $\sim 10^{16}$ G threading the BH horizon are required. These results therefore provide a statistical benchmark for evaluating magnetically powered central engine models in sGRBs. From a theoretical perspective, numerous general relativistic magnetohydrodynamic (GRMHD) simulations of BNS mergers have investigated the emergence of magnetically powered jets resembling the BZ mechanism. These studies generally find that successful jet launching is associated with the presence of a large-scale poloidal magnetic field threading the remnant BH, with characteristic field strengths spanning a broad range of $B \sim 10^{14}$–$10^{17}$\,G. The substantial dispersion in the inferred magnetic field strengths reflects the sensitivity of post-merger magnetization to a variety of modeling choices and physical assumptions, including the initial magnetic field morphology and strength, orbital eccentricity, NS masses, and the adopted nuclear EOS \citep{rgb+11,clk+19,ltq+20,rts+20,rts+21,srs+22,btr+24,hkk+25}. Within this theoretical landscape, our results tend to favor scenarios in which efficient magnetic field amplification leads to the development of relatively strong, ordered poloidal fields near the BH horizon. Conversely, models that systematically produce weaker magnetic fields may face difficulties in accounting for the energetics of sGRBs within the BZ framework, if such jets are assumed to dominate the observed population. Moreover, given the large amplification factors required to bridge the gap between plausible pre-merger NS magnetic fields and the inferred post-merger BH magnetization, our results provide indirect support for the effectiveness of magnetic field amplification through turbulent dynamo processes. These include both post-merger MRI-driven amplification and large-scale $\alpha\Omega$ dynamo action \citep{krs+24}, as well as subgrid prescriptions for unresolved KHI–driven turbulence that assume magnetic field saturation at ultra-strong levels \citep{gzd+15,acr25}. Most recently, \citet{2026arXiv260827555K} showed that merger-driven magnetic field amplification can generate a large-scale field capable of powering a BZ jet whose kinetic energy is compatible with that required by the GRB~170817A afterglow, supporting the ultra-strong post-merger magnetization inferred here. In this sense, observationally inferred constraints on the BH magnetic field provide a valuable complementary perspective to numerical simulations, helping to narrow the physically viable parameter space for post-merger magnetization in BNS systems.

Moreover, in the BZ framework, GRMHD simulations have shown that large-scale poloidal magnetic flux can be efficiently advected inward through the accretion disk and accumulate near the BH horizon, where it eventually reaches a saturated state \citep{tnm+11,mtb+12}. Motivated by this picture, we obtain an order-of-magnitude estimate of the magnetic field strength associated with the disk region supplying the BH-threading magnetic flux by assuming that the large-scale magnetic fluxes threading the BH horizon and the corresponding region of the accretion disk are comparable. Under this assumption, the characteristic disk magnetic field strength can be expressed as $B_\mathrm{disk}\approx(\frac{10~ \mathrm{km}}{R_\mathrm{disk}})^2\times10^{14.1-15.4}$ G, where $R_\mathrm{disk}$ denotes the characteristic radius of the flux-supplying region. Comparing this estimate with the magnetic field strengths typically found in the inner regions of post-merger accretion disks ($\sim10^{15}$\,G; \citealt{kks+18,gcr+25}) suggests that the magnetic flux feeding the BH is predominantly supplied by disk regions at radii of order $\sim10$\,km. This inferred radius is considerably smaller than the characteristic size of post-merger accretion disks ($20-100$ km), while the densest part of the disk is typically located at radii of $\sim30$ km \citep{fm+13,rh+24}. These results therefore suggest that the magnetic flux threading the BH horizon is supplied primarily by the innermost regions of the accretion flow. Such a picture is qualitatively consistent with theoretical studies of magnetic-flux transport, which indicate that the magnetic flux accumulated on the BH is primarily controlled by the transport of magnetic flux through the inner accretion flow \citep{lpp+94,go+12}.

Finally, we emphasize that the amplification factor $\eta_B$ in this work refer to the net enhancement of the magnetic field strength from the pre-merger NS surface to the BH horizon. It therefore encapsulates the cumulative effects of magnetic field amplification and transport throughout the merger and post-merger evolution, rather than the efficiency of any single instability operating at a specific stage. We note that a detailed, time-dependent modeling of the magnetic field evolution during individual mergers is beyond the scope of this work, and future high-resolution GRMHD simulations will be required to further quantify the relative roles of different amplification mechanisms.

\section{Discussion}
\label{sec:discussion}

In this section, we discuss the assumptions and approximations adopted in our analysis. These choices are primarily motivated by the exploratory nature of this work and the goal of connecting the inferred energetics of sGRBs to the properties of their central engines. Specifically, we examine: 
(i) the impact of the possible selection bias introduced, including the effect of adopting a fiducial initial Lorentz factor ($\Gamma_0$= 300);
(ii) the assumption of the cutoff for the post-merger BH mass and the dependence of the results on different EOSs; 
and (iii) the hypothesis that all GRBs in our sample originate from BNS mergers. We assess how these assumptions may affect our results and evaluate the robustness of our main conclusions.

\subsection{Robustness against sample selection and model assumptions}
\label{sub:robustness}

\begin{figure}
\centering
\includegraphics[width=0.99\columnwidth]{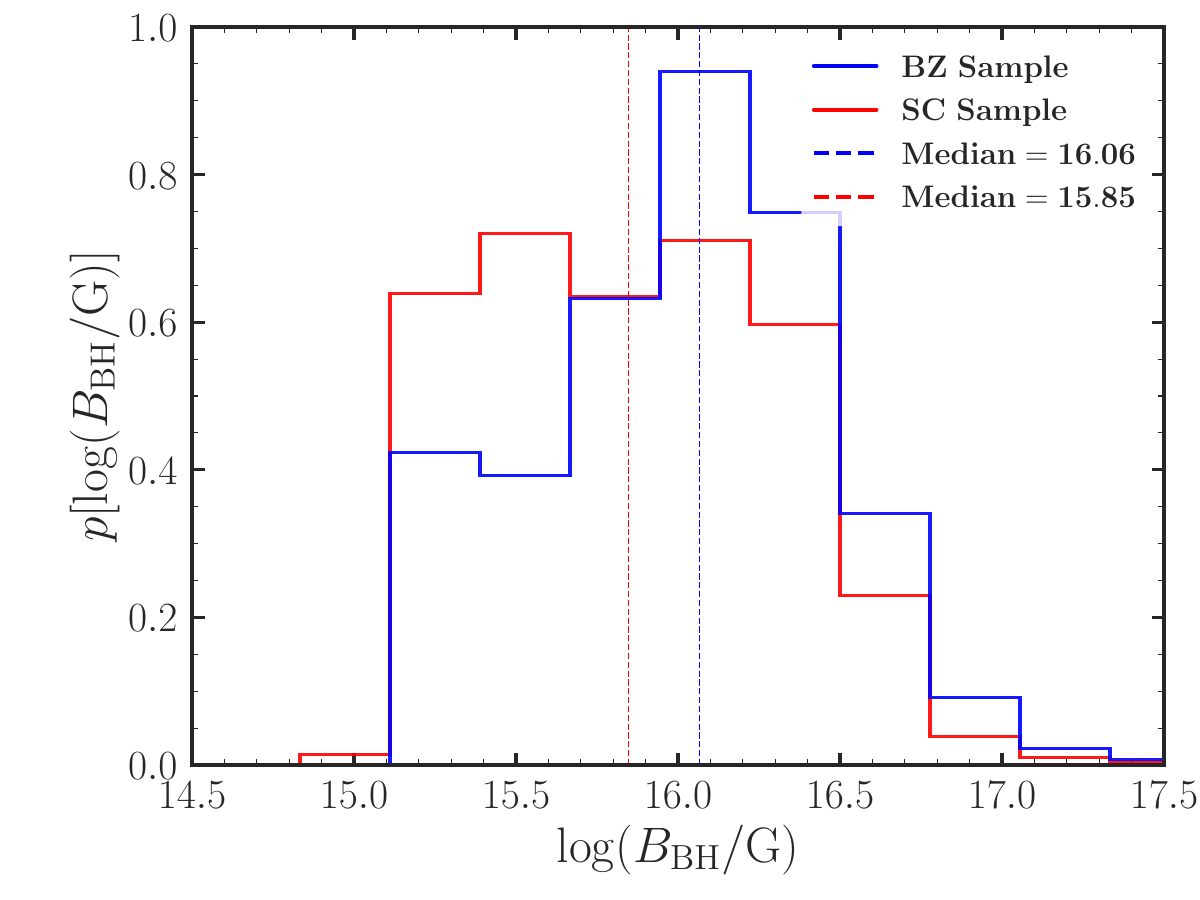}
\caption{ 
Distribution of the inferred BH magnetic field strength for the selected sample and the SC sample. The vertical dashed lines mark the median values of the corresponding distributions.
}
\label{fig:wellvsall}
\end{figure}

We examine the robustness of our inferred BH magnetic field strength against uncertainties associated with sample selection and model assumptions. 
First, we test the impact of potential contamination from alternative central engine processes.
Specifically, we exclude bursts that exhibit X-ray plateau features not accounted for in the FS model listed in \citet{2024A&A...692A..73G}, as such plateaus may indicate additional energy injection from a long-lived central engine \citep[e.g. magnetar spin-down or fallback accretion,][]{knj+08,lz+14,rgd+14}. The resulting magnetic field distribution of this restricted BZ sample, compared to that of the SC sample, is shown in Figure~\ref{fig:wellvsall} for $\tilde{\Lambda}_\mathrm{min}$. The difference between the two distributions corresponds to a shift of less than a factor of $\sim1.7$, indicating only a minor impact.

\begin{figure}
\centering
\includegraphics[width=0.99\columnwidth]{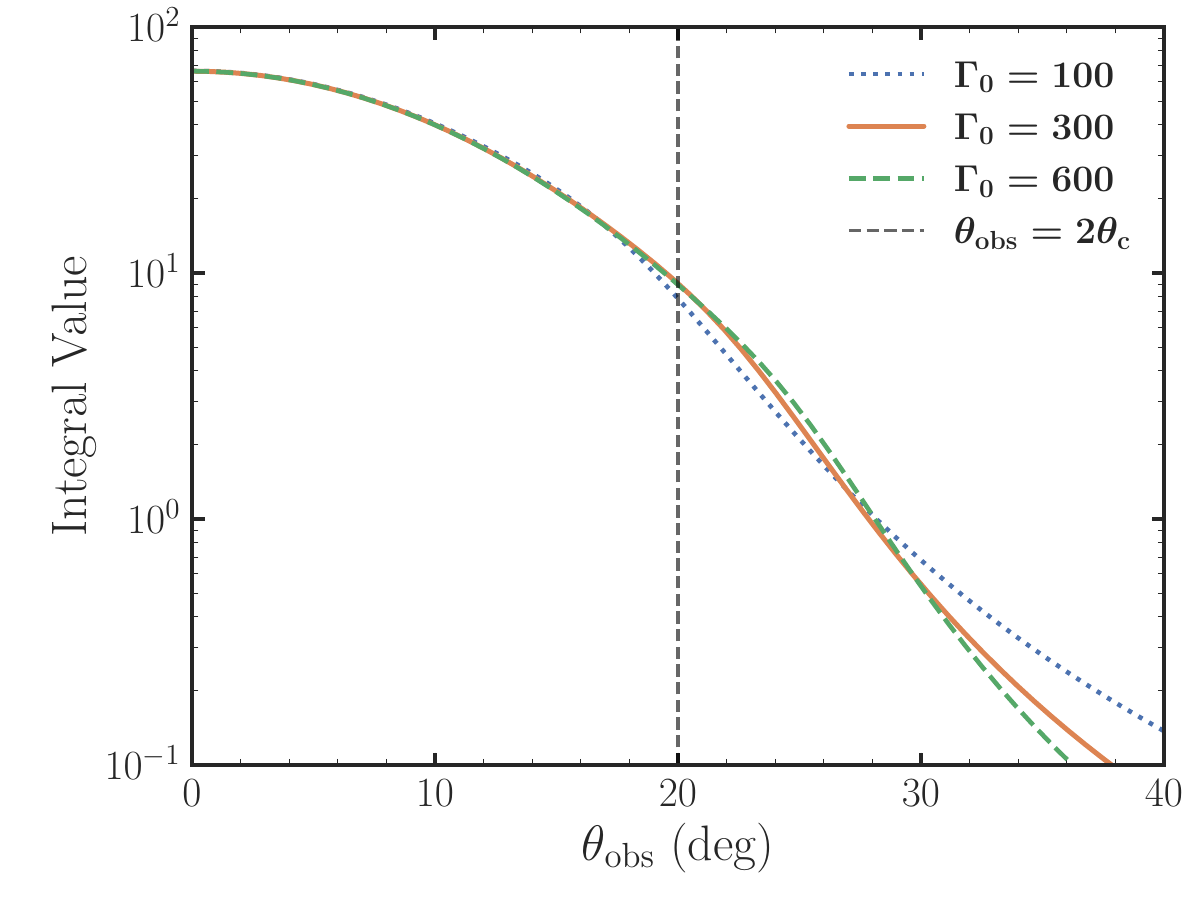}
\caption{Dependence of the integral conversion term in Equation~(\ref{con:egammatoiso}) on the viewing angle for different initial Lorentz factors $\Gamma_0$. For all curves, the jet core angle $\theta_\mathrm{c}$ and the truncation angle $\theta_\mathrm{w}$  are fixed at 0.1 and 0.4, respectively, while different values of $\Gamma_0$ are indicated in the legend. The vertical dashed line marks a viewing angle equal to twice the jet core angle, below which a large fraction of the bursts in our sample are located.}
\label{fig:lorentz}
\end{figure}

We further explore the dependence of our results on the assumed initial Lorentz factor. In Section~\ref{sub:energetics}, we adopt a fiducial value $\Gamma_0 = 300$ to estimate the $\gamma$-ray radiative efficiency $\eta_\gamma$. Figure~\ref{fig:lorentz} shows the dependence of the integral term appearing in the conversion between the isotropic-equivalent energy and the true energy on the viewing angle for several values of $\Gamma_0$. For all curves, the jet core angle $\theta_\mathrm{c}$ and the truncation angle $\theta_\mathrm{w}$ are fixed at 0.1 and 0.4, respectively, while the vertical dashed line marks a viewing angle $\theta_\mathrm{obs}=2\theta_\mathrm{c}$, below which a large fraction of bursts in our sample are located. As shown in the figure, the integral term exhibits only a weak dependence on $\Gamma_0$ over the relevant range of viewing angles. Even at $\theta_\mathrm{obs}\simeq2\theta_\mathrm{c}$, varying $\Gamma_0$ from 100 to 600 induces only a modest change. Quantitatively, the median value of $\eta_\gamma$ varies from $\simeq0.29$ for $\Gamma_0=100$ to $\simeq0.38$ for $\Gamma_0=600$, compared to $\simeq0.35$ for our fiducial choice $\Gamma_0=300$. These variations are small relative to the overall dispersion of $\eta_\gamma$ in our sample and have a negligible impact on the inferred distribution of $B_{\rm BH}$. We therefore conclude that our main result—the requirement of magnetic fields at the level of $\sim10^{16}~\mathrm{G}$ if the BZ mechanism powers sGRBs—is insensitive to both sample selection effects and reasonable variations in the initial Lorentz factor.

\subsection{BH formation criterion and mass threshold}
\label{sub:mass threshold}

\begin{figure}
\centering
\includegraphics[width=0.99\columnwidth]{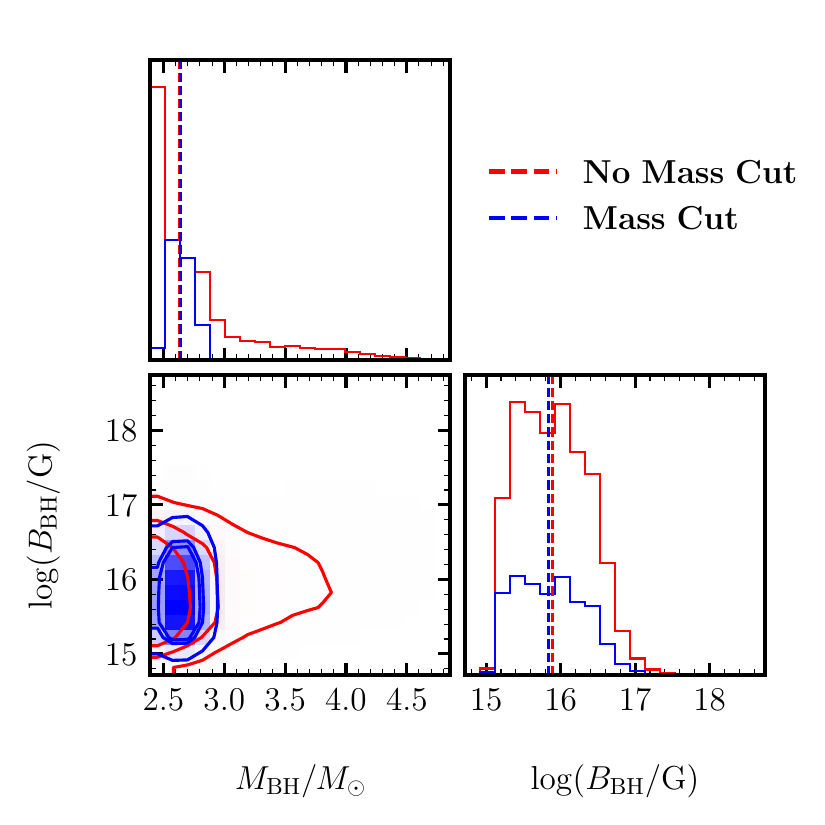}
\includegraphics[width=0.99\columnwidth]{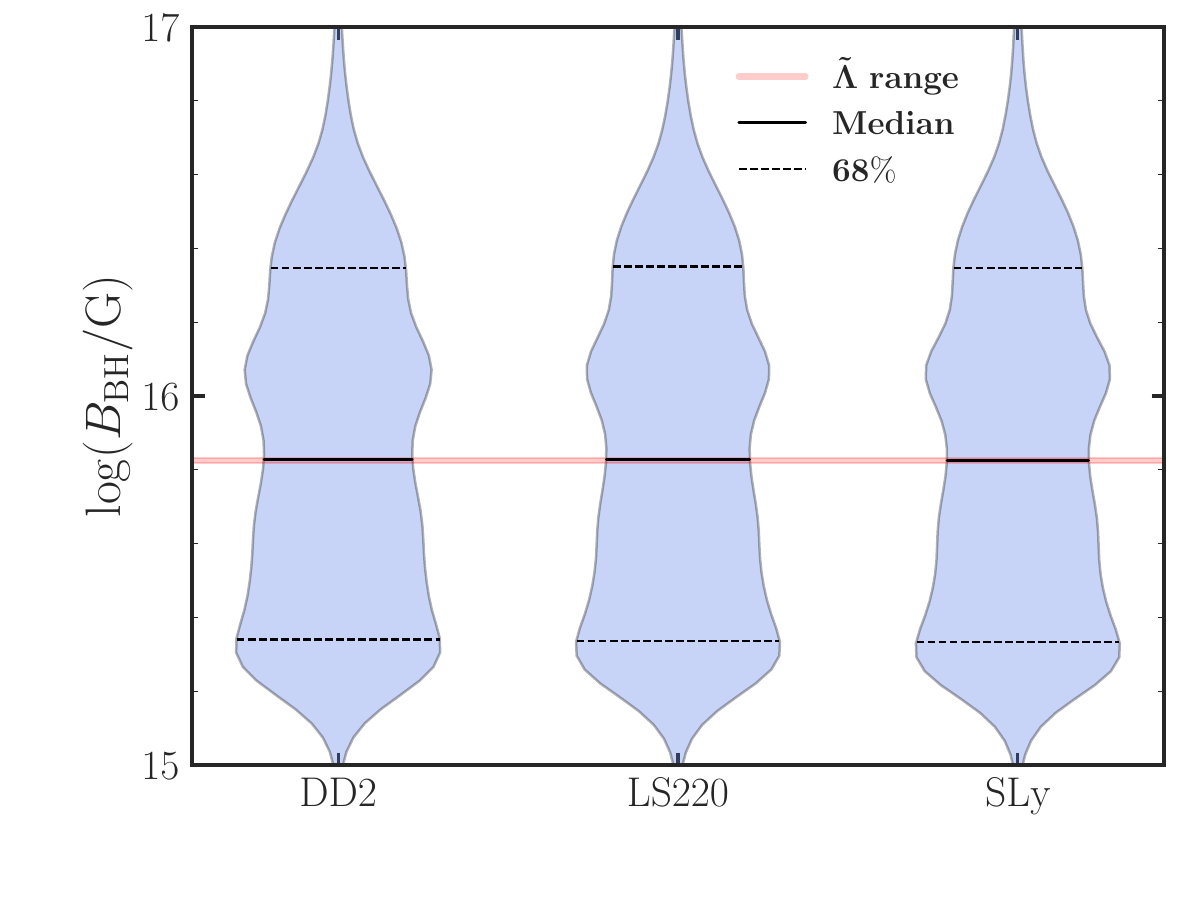}
\caption{Top:Marginal distributions of the BH magnetic field $\log(B_\mathrm{BH})$ and the BH mass $M_\mathrm{BH}$, together with their joint distribution for $\tilde{\Lambda}_\mathrm{min}$.
The red and blue contours correspond to the cases without imposing a limit on the BH mass and with a mass threshold, respectively.
The vertical dashed lines indicate the median values of the corresponding marginal distributions. We find that imposing a BH mass cut leads to only a modest shift in the inferred magnetic field distribution, which remains centered at $B_{\mathrm{BH}}\sim10^{16}~\mathrm{G}$. Bottom: Inferred BH magnetic field strength for three EOSs. The red shaded region indicates the range of median values derived from the $\tilde{\Lambda}_\mathrm{max}$ and $\tilde{\Lambda}_\mathrm{min}$ cases. The blue violin plots show the posterior distributions for each EOS. The black solid lines mark the median values, while the black dashed lines denote the 68\% credible intervals.
}
\label{fig:B_M}
\end{figure}

To restrict the synthetic population to systems that are likely to form BHs while retaining sufficiently massive disks to power relativistic jets, we adopt a fiducial mass range of $2.5<M_{\mathrm{BH}}/M_\odot<2.8$ in Section~\ref{sub:BHmass}.
In this subsection, we examine the sensitivity of our results to this choice. In particular, we explore how relaxing the mass threshold affects the inferred distributions of the BH mass and magnetic field, and assess how the inferred magnetic field changes when adopting different EOSs together with the corresponding collapse thresholds.

Although adopting $M_{\mathrm{BH}}>2.5M_\odot$ selects remnants that are expected to collapse into BHs rather than remain stable NSs \citep{mjm+22}, BH formation alone does not guarantee the presence of a sufficiently massive torus capable of powering an sGRB jet. For a given EOS and binary mass ratio, systems with a total mass above the prompt-collapse threshold ($M_{\mathrm{thr}}$), collapse directly into a BH and generally leave a substantially less massive accretion disk, whereas systems below this threshold can form a transient differentially rotating remnant whose delayed collapse is more favorable for producing a BH--torus system and launching a relativistic jet. NR calculations show that $M_{\mathrm{thr}}$ depends on both the EOS and the binary mass ratio, with values near $2.8M_\odot$ representing the lower end of the predicted threshold range for the EOSs relevant to this work \citep{2020PhRvL.125n1103B,rts+21}. To reduce contamination from systems likely to undergo prompt collapse, we therefore additionally require $M_{\mathrm{BH}}<2.8M_\odot$. To assess the sensitivity of our results to the adopted BH mass threshold, the top panel of Fig.~\ref{fig:B_M} presents the marginal distributions of the BH magnetic field $\log(B_\mathrm{BH})$ and BH mass $M_\mathrm{BH}$, together with their joint distribution, for the case of $\tilde{\Lambda}_\mathrm{min}$. We compare the results obtained with and without imposing a mass cut. As expected, introducing a mass threshold significantly alters the marginal distribution of $M_\mathrm{BH}$. However, the joint distributions in the $M_\mathrm{BH}$--$\log(B_\mathrm{BH})$ plane remain qualitatively similar in the two cases. The primary effect of the mass cut is a translation of the joint distribution, while its overall shape and correlation structure are largely preserved.
Consequently, the inferred distribution of the BH magnetic field is only weakly affected by the choice of the mass threshold. Quantitatively, the median value of $B_\mathrm{BH}$ changes only slightly, from $7.6\times10^{15}$ G to $7.0\times10^{15}$ G, indicating that our conclusions regarding the characteristic magnetic field strength at the BH horizon are robust against reasonable variations in the assumed NS mass. We further test the sensitivity of our results to the assumed BNS mass distribution by adopting the mass spectrum inferred from Galactic BNS observations reported in ATNF catalog \citep{mht+05}. Even under this significantly different mass distribution, the resulting change in the inferred magnetic field strength remains below $\sim20\%$. This further supports the robustness of our results against uncertainties in the underlying mass spectrum.

In addition to the BH mass threshold, the inferred magnetic field may also depend on the assumed NS EOS. In our main analysis, this uncertainty has already been bracketed by adopting the minimum and maximum values of the tidal deformability ($\tilde{\Lambda}_\mathrm{min}$ and $\tilde{\Lambda}_\mathrm{max}$). To further illustrate the robustness of this treatment, we additionally compute the results using several representative EOSs and compare the corresponding magnetic field values with the range obtained in our fiducial analysis. Specifically, we consider the SLy, DD2 and LS220 EOSs \citep{ls+91,dh+01,trk+10}, which are all consistent with current observational constraints. For each EOS, the calculations of the tidal deformability $\tilde{\Lambda}$ and the corresponding NS mass threshold follow the same procedure as described in \citet{zlc+23}. The results are shown in the bottom panel of Figure~\ref{fig:B_M}. We find that the median magnetic field obtained for each specific EOS lies within the range defined by the $\tilde{\Lambda}_\mathrm{min}$ and $\tilde{\Lambda}_\mathrm{max}$ cases, further confirming the robustness of the inferred BH magnetic field distribution.

\subsection{Progenitor systems: BNS versus NSBH mergers}
\label{sub:BNS NSBH}

Our main results rely on the assumption that sGRBs predominantly originate from BNS mergers. The joint detection of the GW event GW170817 and its associated sGRB GRB~170817A has provided direct and compelling observational evidence that BNS mergers can produce sGRBs \citep{aaa+17}. Nevertheless, NSBH mergers may also contribute to the sGRB population. From the GW perspective, the LIGO--Virgo--KAGRA Collaboration has confidently detected the NSBH merger GW200115 and reported another candidate event, GW200105, providing direct evidence for the existence of such binaries in nature \citep{aaaa+21}. In principle, these systems are also capable of producing sGRBs under suitable physical conditions.

However, current observational evidence generally favors a BNS-dominated origin for sGRBs. Analyses of the GWTC-4 catalog infer an NSBH merger rate of $\sim6.7$–$32.8\,\mathrm{Gpc^{-3}\,yr^{-1}}$, compared with a BNS merger rate of $\sim5.1$–$154.7\,\mathrm{Gpc^{-3}\,yr^{-1}}$ \citep{ttt+25,2026arXiv260527226T}. Independent support also comes from statistical studies of the sGRB population. For example, \citet{berger+11} showed that the observed offset distribution of sGRBs relative to their host galaxy centers is in good agreement with theoretical expectations for BNS mergers. In addition, \citet{nfd+22} demonstrated that the stellar mass–metallicity relation of sGRB host galaxies is inconsistent with the scenario in which all sGRBs originate from NSBH mergers. Overall, these observational results suggest that the majority of sGRBs are likely produced by BNS mergers.

Moreover, from the theoretical perspective, not all NSBH mergers are expected to produce observable electromagnetic counterparts. The formation of a relativistic jet requires that the NS be tidally disrupted outside the BH innermost stable circular orbit so that a sufficiently massive accretion disk can form. This condition is only satisfied within a limited region of parameter space, typically requiring relatively high (prograde) BH spins and NSs with sufficiently large radii \citep{bpt+72}. As a consequence, a large fraction of NSBH mergers are expected to remain electromagnetically dark. Population-synthesis studies further suggest that only a modest fraction of NSBH mergers are capable of producing detectable sGRBs under realistic assumptions about BH spin distributions and NS EOS \citep{f+21}.

Taken together, these considerations suggest that, although NSBH mergers may contribute to the observed sGRB population, their overall contribution is likely to be subdominant. Therefore, we expect that only a small fraction of the GRBs in our sample, if any, originate from NSBH mergers, and that our main conclusions (based on the assumption that sGRBs predominantly arise from BNS mergers) are unlikely to be significantly biased by the contamination of NSBH mergers in the sample. 

\section{Conclusions}
\label{sec:conclusion}

In this paper, we compile a broadband afterglow dataset of 40 sGRBs and model their multi-wavelength light curves within the framework of an off-axis Gaussian jet. According to the model fittings, we derive the distributions of key energetic parameters of sGRBs and characterize them using parametric functions. By combining the inferred energetics with a synthetic compact-binary population model and assuming jet launching via the BZ process, we further estimate the distribution of the magnetic field strength at the BH horizon. We summarize our main conclusions below:
\begin{enumerate}
    \item 
    By modeling the broadband afterglows of 15 SC sGRBs, we investigate the statistical properties of their jet energetics. We find median values of $E_0 \simeq 3.5\times10^{52}~\mathrm{erg}$, $E_\mathrm{k} \simeq 3.8\times10^{49}~\mathrm{erg}$, and a radiative efficiency of $\eta_\gamma \simeq 0.35$. The inferred distributions of $E_0$, $E_\mathrm{k}$, and $\eta_\gamma$ can be characterized by analytical functions, providing observational constraints on the energetics and radiative properties of sGRB jets.
    
    \item Combining the inferred jet energetics with synthetic BNS populations, NR fitting formulae, and the assumption of BZ-powered jets, 
    we infer a characteristic accretion-disk-supported magnetic field strength of $B_\mathrm{BH}\sim10^{16}$ G threading the BH horizon.
    Comparing these values with the magnetic field distribution of pre-merger NSs further implies a characteristic magnetic field amplification factor of $\eta_B\sim10^3$ during the merger and post-merger evolution, providing an independent observational constraint on magnetic field amplification in BNS mergers.
    \item 
    Extending the analysis to the full sample introduces more pronounced high-energy and high-field tails, predominantly contributed by the less well-constrained bursts. The median inferred magnetic field increases from approximately $7\times10^{15}$ G for the SC sample to $2\times10^{16}$ G for the full sample. Although the inferred median values and distribution shapes are sensitive to sample composition, both samples consistently indicate a characteristic magnetic field strength of order $10^{16}$ G within the BZ framework. 
\end{enumerate}
However, due to the limited sample of well-characterized sGRB afterglows, as well as the incomplete understanding of selection effects and the magnetic field geometry following BNS mergers, our constraints on the sGRB energetics and the inferred magnetic field strength remain subject to substantial uncertainties and potential systematic biases. Nevertheless, future joint electromagnetic and GW observations, together with improved population synthesis models and more comprehensive afterglow monitoring, will enable more stringent tests of this framework and help to better characterize the diversity of sGRB central engines. If future observations continue to support such a high magnetic field strength, this may indicate that, within the framework in which sGRBs are powered by relativistic jets launched via the BZ process in BNS mergers, the required level of magnetization favors scenarios in which the post-merger remnant undergoes efficient magnetic field amplification, for example through turbulence-driven processes and saturation mechanisms.
\section*{Data availability}
The full versions of Tables \ref{tab:xdata}, \ref{tab:otable}, and \ref{tab:rtable} are
available in electronic form at the CDS via
anonymous ftp to cdsarc.u-strasbg.fr
or via
\url{http://cdsweb.u-strasbg.fr/cgi-bin/qcat?J/A+A/}.
\begin{acknowledgements} 
We thank Professor Yanan Wang and Professor Junqiang Ge for their insightful discussions and suggestions.   
This work is partly supported by the National Natural Science Foundation of China (grant nos. 12533009, and 12273050), the Strategic Priority Program of the Chinese Academy of Sciences (grant no. XDB0550303), the National Astronomical Observatory of China (grant no. E4TG660101), and the Postdoctoral Fellowship Program of CPSF under Grant Number GZB20250735 (ZC).
\end{acknowledgements}

\bibliographystyle{aa} 
\bibliography{reference} 

\clearpage

\begin{appendix}

\section{$k$-correction}
\label{appendix:k-correction}

\begin{table}
\caption{Prompt-emission spectral parameters adopted for the sGRB
$k$-corrections.}
\label{tab:k-correction}
\centering
\begin{tabular}{lccccc}
\hline\hline
\multirow{2}{*}{GRB} & $E_\mathrm{peak}$ & \multirow{2}{*} {$\alpha$} & \multirow{2}{*} {$\beta$} & \multirow{2}{*} {$k$} & \multirow{2}{*} {Reference}\\
    &   (keV)       &      &   &  &\\
\hline
211227A & 192 & -1.34 & -2.26&3.11 & (1) \\
211211A & 647 & -1.3 & -2.4&5.20 & (2) \\
210323A & 2100 & -0.97 & -3.02&16.01 & (3) \\
200219A & 1400 & -0.6 & - &20.12& (4) \\
180727A & 79.2 & -1.54 & -2.46&2.34 & (5) \\
170428A & 982 & -0.47 & -2.46 &19.97& (6) \\
150120A & 130 & -1.43 & - &2.01& (7) \\
130603B & 660 & -0.73 & - &6.95& (8) \\
120804A & 135 & -0.89 & - &1.73& (9) \\
101224A & 330 & -0.83 & - &3.28& (10) \\
100206A & 506 & -0.18 & - &7.52& (11) \\
100117A & 287 & -0.14 & - &3.48&(12) \\
090510 & 4400 & -0.8 & -2.6 &38.65& (13) \\
071227 & 1000 & -0.7 & - &11.54& (14) \\
070714B & 1120 & -0.86 & -&10.63 & (15) \\
061006 & 966 & -0.75 & - &10.44& (16) \\
060614 & 302 & -1.57 & - &2.83& (17) \\
051221A & 402 & -1.08 & - &3.52& (18) \\
\hline
\end{tabular}
\tablebib{
(1) \citet{2022GCN.31544....1T};
(2) \citet{2021GCN.31210....1M};
(3) \citet{2021GCN.29709....1H};
(4) \citet{2020GCN.27137....1B};
(5) \citet{2018GCN.23053....1V};
(6) \citet{2017GCN.21045....1T};
(7) \citet{2015GCN.17319....1V};
(8) \citet{2013GCN.14771....1G};
(9) \citet{2012GCN.13614....1S};
(10) \citet{2010GCN.11489....1M};
(11) \citet{2010GCN.10381....1V};
(12) \citet{2010GCN.10345....1P};
(13) \citet{2009GCN..9336....1G};
(14)\citet{2007GCN..7155....1G};
(15) \citet{2007GCN..6638....1O};
(16) \citet{2006GCN..5717....1U};
(17) \citet{2006GCN..5264....1G};
(18) \citet{2005GCN..4394....1G}.
}
\tablefoot{
The reported $E_{\rm peak}$ values are measured in the
observer frame. We adopt the Band function when both $\alpha$ and
$\beta$ are available \citep{1993ApJ...413..281B}, and a cutoff power-law model otherwise. The
$k$-correction converts the observed 15--150~keV fluence to the
1--$10^{4}$~keV rest frame energy band.
}
\end{table}
Table~\ref{tab:k-correction} lists the prompt emission spectral parameters adopted for the burst-by-burst $k$-corrections, together with the resulting correction factors. These values are used to convert the observed 15--150~keV fluences to the 1--$10^{4}$~keV rest frame energy band.

\section{Afterglow model validation}
\label{appendix:vegas}

To assess whether the treatment of the early blast-wave dynamics in \texttt{afterglowpy} affects our results, we performed an independent analysis using \texttt{VegasAfterglow} through \texttt{Redback} \citep{2024MNRAS.531.1203S}. \texttt{VegasAfterglow} explicitly follows the coasting-to-deceleration transition and therefore provides a more detailed description of the early radial evolution. Its treatment of lateral spreading, however, relies on a prescribed smooth transition function, such that its more detailed early radial dynamics does not necessarily imply uniformly greater accuracy throughout the subsequent jet evolution. Nevertheless, the two codes produce broadly similar light curve behavior over their common regimes of applicability \citep{2026JHEAp..5000490W}. We therefore use \texttt{VegasAfterglow} as a complementary and independent test of the whole time dynamics. To ensure a controlled comparison, we adopted the same truncated Gaussian jet structure as in our \texttt{afterglowpy} analysis.

We first refitted the 15 SC bursts after including the early-time multi-wavelength observations. We found that these observations still cannot be consistently reproduced by the standard FS model. Their exclusion is therefore unlikely to result from the approximation of a coasting phase in \texttt{afterglowpy}, and instead indicates the presence of additional emission components not included in the standard FS model.

\begin{figure}
\centering
\includegraphics[width=0.99\columnwidth]{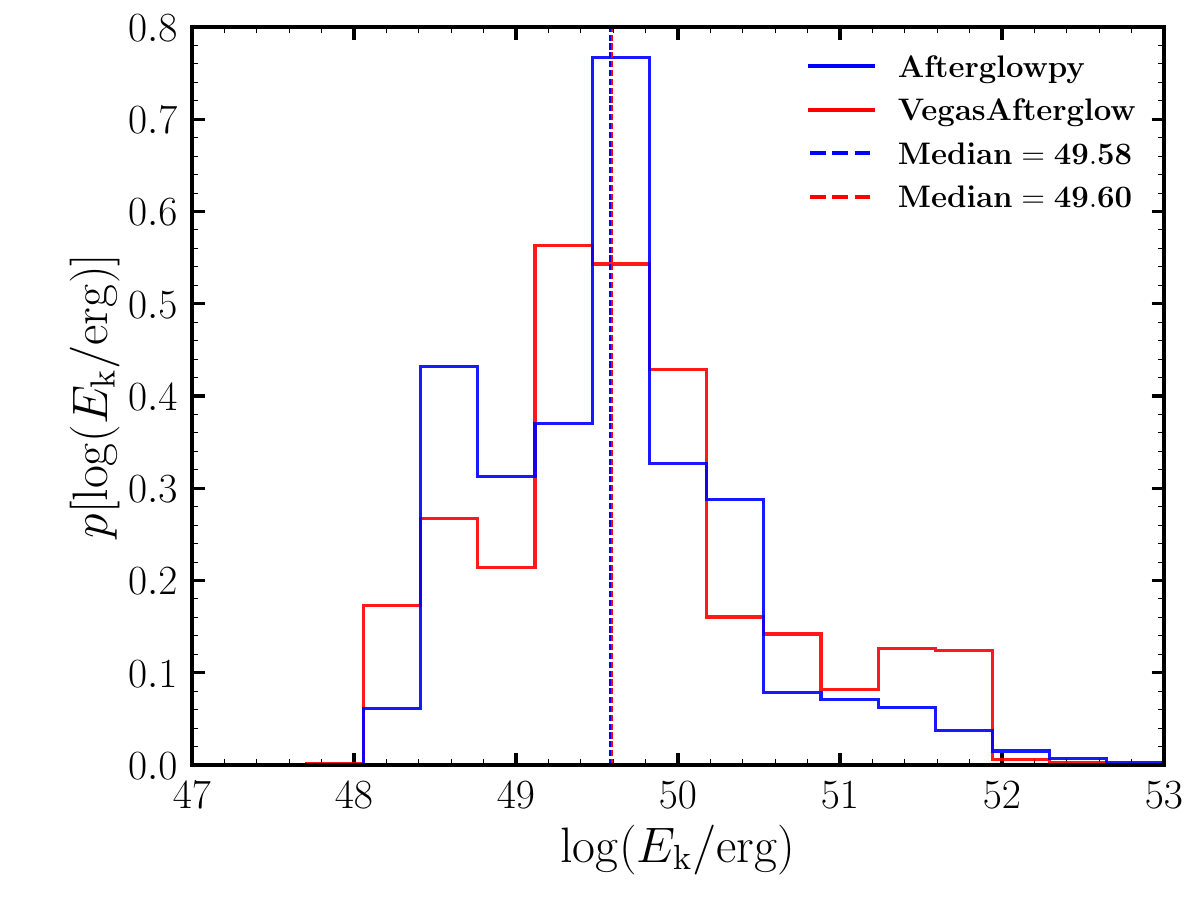}
\caption{Distributions of the kinetic energy $E_{\mathrm{k}}$ inferred for the 15 SC bursts using \texttt{afterglowpy} (blue) and \texttt{VegasAfterglow} (red). The dashed vertical lines mark the corresponding median values, $\log(E_{\mathrm{k}}/\mathrm{erg}) = 49.58$ and $49.60$. The agreement between the two distributions indicates that the
treatment of the early blast-wave dynamics has a negligible effect on
the inferred kinetic energies.
}
\label{fig:vegas_energy}
\end{figure}

We then refitted the same 15 SC bursts using only the data adopted in our original analysis. Figure~\ref{fig:vegas_energy} compares the distributions of the kinetic energy inferred with \texttt{afterglowpy} and \texttt{VegasAfterglow}. The two distributions are consistent, with median values of $\log(E_{\mathrm{k}}/\mathrm{erg})=49.58$ and $49.60$, respectively. Thus, although the two codes differ in their treatment of the blast-wave evolution, this difference has a negligible effect on the kinetic energy estimates relevant to our analysis. This direct empirical comparison supports both the robustness of the energetic results obtained with \texttt{afterglowpy} and its use as the baseline model for our homogeneous population analysis.

\onecolumn

\section{Fitting results and broadband data}

This appendix presents the complete broadband afterglow modeling results and observational data used in this work.

\begin{figure*}[ht!]
\centering
\includegraphics[width=\textwidth,height=\textheight,keepaspectratio]{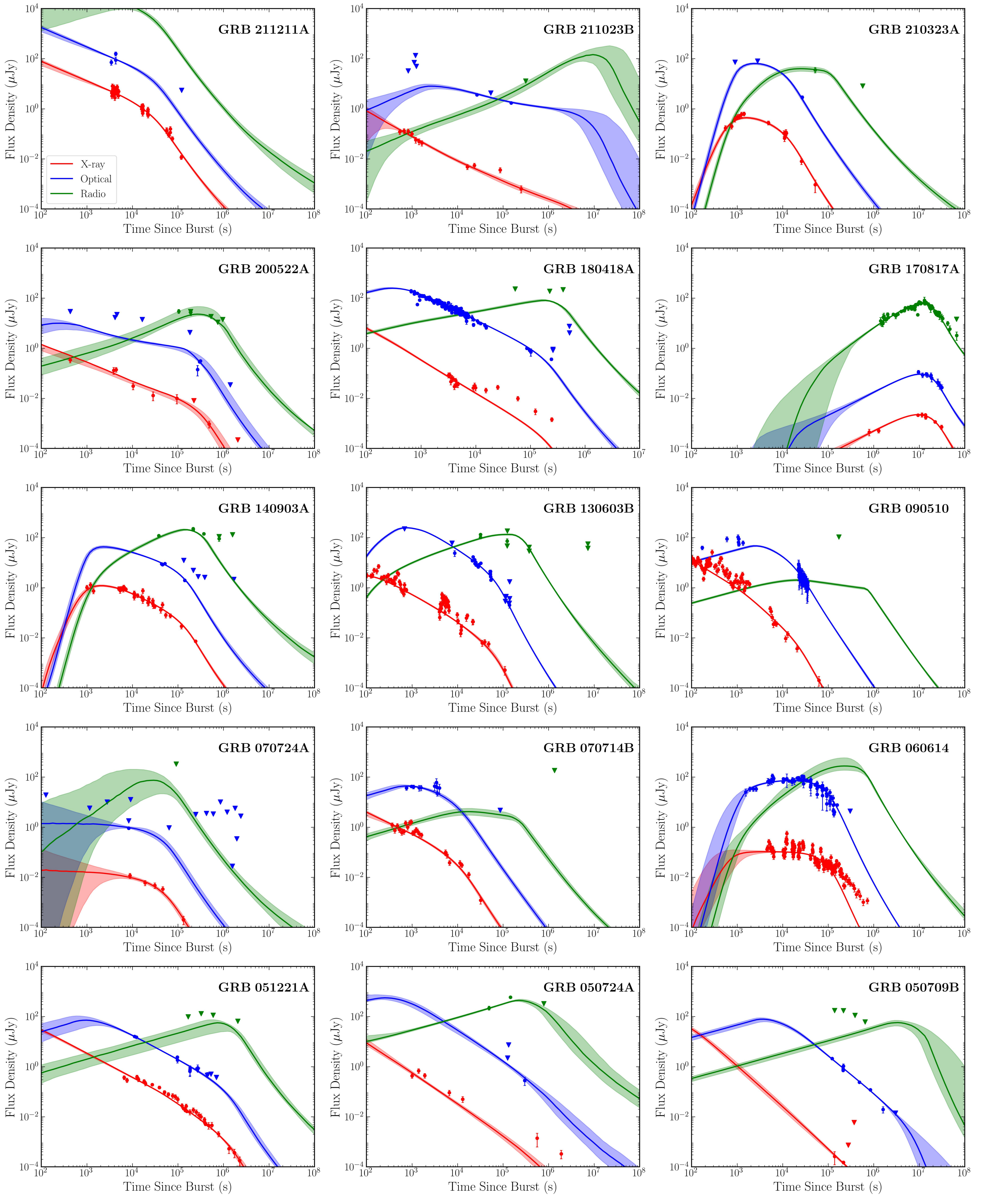}
\caption{Broadband afterglow fits for the 15 SC GRBs. Red, blue, and green points denote X-ray, optical, and radio observations; triangles mark upper limits. Solid lines show median model light curves generated from hundreds of posterior samples, with shaded regions indicating the 16th–84th percentile uncertainties. Optical and radio fluxes at different bands are converted to those at $5\times10^{14}$ Hz and $6\times10^{9}$ Hz, respectively, for visual clarity.
\label{fig:well}}
\end{figure*}

\begin{figure*}[t]
\centering
\includegraphics[width=\textwidth,height=\textheight,keepaspectratio]{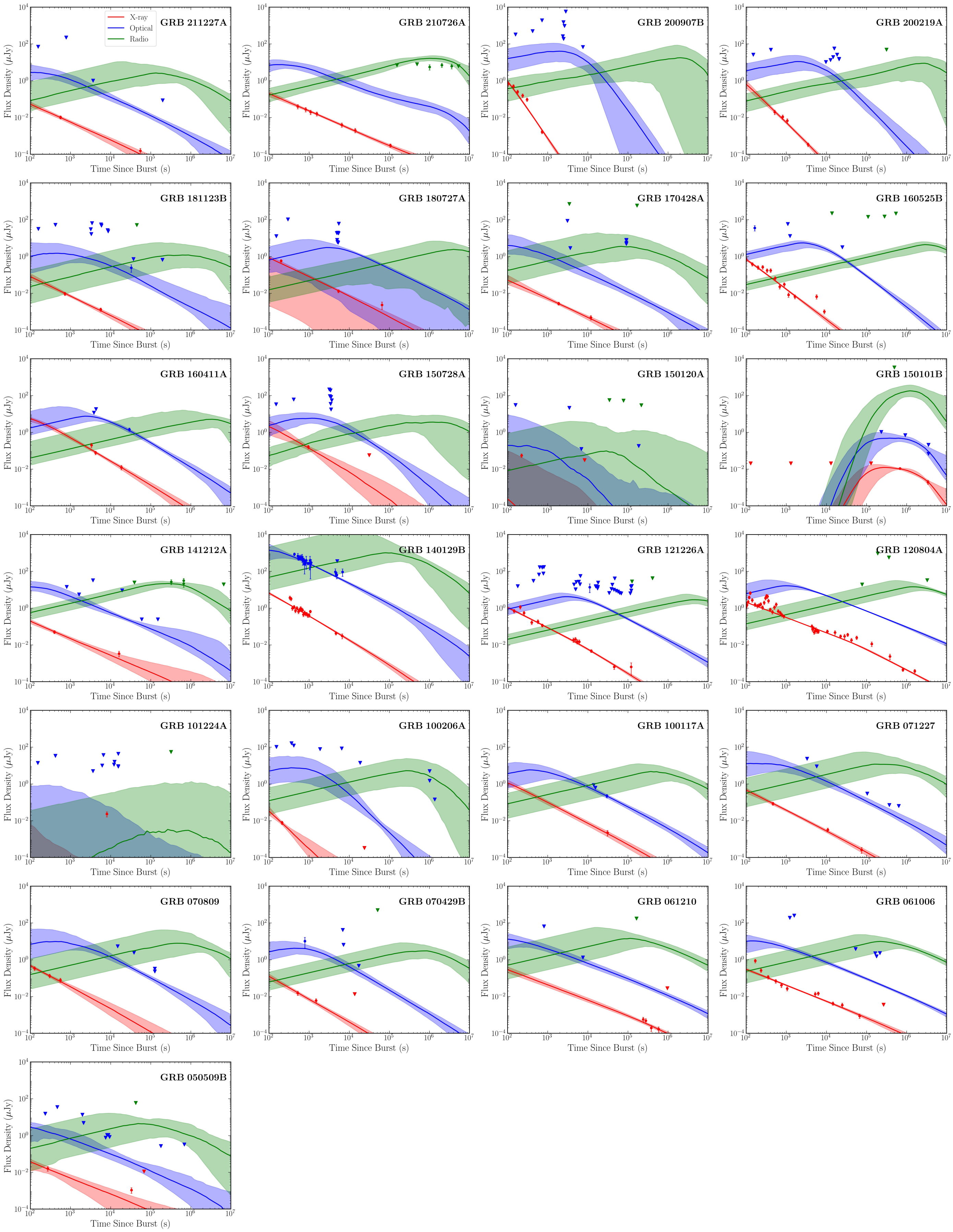}
\caption{
Same as Fig.~\ref{fig:well}, but for the 25 non-SC GRBs.
\label{fig:others}}
\end{figure*}

Tables \ref{tab:well} and \ref{tab:other} summarize the posterior median values and 68\% credible intervals of the Gaussian jet model parameters for the 15 SC GRBs and the remaining 25 GRBs, respectively. The corresponding broadband afterglow fits are shown in Figures \ref{fig:well} and \ref{fig:others}. 

Tables \ref{tab:xdata}, \ref{tab:otable}, and \ref{tab:rtable} provide the complete X-ray, optical/NIR, and radio afterglow catalogs compiled in this work.\footnote{Machine-readable versions of the data tables are available at \href{https://github.com/zeng-cw/afterglow-data}{https://github.com/zeng-cw/afterglow-data}.}

\begin{table*}
\caption{Fitting results of broadband modeling of strictly constrained GRB light curves given a Gaussian jet model.}
\label{tab:well}
\centering
\begin{tabular}{lcccccccc}
\hline\hline
\multirow{2}{*}{GRB} & $\theta_{\mathrm{w}}$ & $\theta_{\mathrm{obs}}$ & $\mathrm{log}~E_0$ &$\theta_{\mathrm{c}}$ & $\mathrm{log}~n_0$ &\multirow{2}{*}{ $p$}&\multirow{2}{*}{ $\mathrm{log}~\epsilon_{\mathrm{e}}$}& \multirow{2}{*}{$\mathrm{log}~\epsilon_{\mathrm{b}}$} \\
    &  (rad) &(rad)  &(erg) &(rad) &($\mathrm{cm^{-3}}$)&  &  & \\
\hline
211211A & $0.090_{-0.037}^{+0.069}$ & $0.034_{-0.020}^{+0.021} $ & $52.72_{-0.60}^{+0.81} $ &$0.038_{-0.013}^{+0.017} $ & $0.05_{-1.11}^{+0.69} $ & $2.01_{-0.01}^{+0.01} $\tablefootmark{a} & $-1.16_{-0.84}^{+0.48} $ & $-3.95_{-0.41}^{+0.90} $\\[4pt]
211023B & $0.115_{-0.033}^{+0.042}$ & $0.101_{-0.029}^{+0.044} $ & $54.42_{-0.58}^{+0.36} $ &$0.030_{-0.009}^{+0.013} $ & $-3.63_{-0.83}^{+0.63} $ & $2.61_{-0.34}^{+0.07} $ & $-0.61_{-0.16}^{+0.08} $ & $-2.02_{-0.55}^{+0.91} $\\[4pt]
210323A & $0.016_{-0.003}^{+0.004}$ & $0.029_{-0.006}^{+0.005} $ & $53.48_{-0.11}^{+0.13} $ &$0.014_{-0.003}^{+0.003} $ & $-0.05_{-0.14}^{+0.51} $ & $2.17_{-0.04}^{+0.05} $ & $-0.60_{-0.12}^{+0.07} $ & $-4.37_{-0.37}^{+0.40} $\\[4pt]
200522A & $0.087_{-0.022}^{+0.030}$ & $0.068_{-0.013}^{+0.015} $ & $52.20_{-0.16}^{+0.16} $ &$0.025_{-0.005}^{+0.006} $ & $-2.78_{-0.61}^{+0.62} $ & $2.08_{-0.03}^{+0.04} $ & $-0.58_{-0.12}^{+0.06} $ & $-1.11_{-0.41}^{+0.39} $\\[4pt]
180418A & $0.076_{-0.017}^{+0.010}$ & $0.065_{-0.015}^{+0.008} $ & $52.97_{-0.11}^{+0.18} $ &$0.069_{-0.017}^{+0.011} $ & $0.58_{-0.73}^{+0.31} $ & $2.26_{-0.04}^{+0.03} $ & $-0.81_{-0.16}^{+0.07} $ & $-4.11_{-0.20}^{+0.40} $\\[4pt]
170817A & $0.311_{-0.040}^{+0.040}$ & $0.402_{-0.032}^{+0.028} $ & $50.88_{-0.05}^{+0.08} $ &$0.070_{-0.012}^{+0.005} $ & $-4.76_{-0.21}^{+0.17} $ & $2.15_{-0.01}^{+0.01} $ & $-0.52_{-0.03}^{+0.01} $ & $-0.61_{-0.14}^{+0.08} $\\[4pt]
140903A & $0.067_{-0.022}^{+0.017}$ & $0.086_{-0.029}^{+0.020} $ & $52.46_{-0.13}^{+0.30} $ &$0.050_{-0.016}^{+0.010} $ & $-0.42_{-1.12}^{+0.61} $ & $2.17_{-0.04}^{+0.04} $ & $-0.66_{-0.24}^{+0.12} $ & $-3.24_{-0.32}^{+0.73} $\\[4pt]
130603B & $0.056_{-0.015}^{+0.011}$ & $0.063_{-0.017}^{+0.013} $ & $52.66_{-0.20}^{+0.27} $ &$0.055_{-0.014}^{+0.011} $ & $0.18_{-0.75}^{+0.45} $ & $2.56_{-0.02}^{+0.02} $ & $-0.63_{-0.15}^{+0.09} $ & $-4.68_{-0.22}^{+0.38} $\\[4pt]
090510 & $0.018_{-0.004}^{+0.003}$ & $0.008_{-0.002}^{+0.002} $ & $52.37_{-0.10}^{+0.15} $ &$0.016_{-0.003}^{+0.004} $ & $-3.18_{-0.78}^{+0.46} $ & $2.44_{-0.03}^{+0.04} $ & $-0.62_{-0.16}^{+0.09} $ & $-1.90_{-0.28}^{+0.47} $\\[4pt]
070724A & $0.054_{-0.025}^{+0.040}$ & $0.053_{-0.021}^{+0.026} $ & $52.95_{-0.40}^{+0.70} $ &$0.022_{-0.008}^{+0.011} $ & $-0.46_{-1.39}^{+0.91} $ & $2.07_{-0.04}^{+0.05} $ & $-0.92_{-0.69}^{+0.33} $ & $-4.44_{-0.41}^{+0.91} $\\[4pt]
070714B & $0.020_{-0.006}^{+0.022}$ & $0.015_{-0.012}^{+0.004} $ & $52.79_{-0.37}^{+0.21} $ &$0.012_{-0.002}^{+0.004} $ & $-1.86_{-0.81}^{+0.67} $ & $2.16_{-0.05}^{+0.06} $ & $-0.57_{-0.12}^{+0.07} $ & $-3.01_{-0.43}^{+0.54} $\\[4pt]
060614 & $0.040_{-0.008}^{+0.012}$ & $0.051_{-0.010}^{+0.017} $ & $53.25_{-0.11}^{+0.20} $ &$0.014_{-0.003}^{+0.004} $ & $-2.22_{-0.58}^{+0.98} $ & $2.57_{-0.02}^{+0.01} $ & $-0.58_{-0.10}^{+0.06} $ & $-3.37_{-0.43}^{+0.41} $\\[4pt]
051221A & $0.190_{-0.074}^{+0.117}$ & $0.063_{-0.038}^{+0.043} $ & $52.20_{-0.34}^{+0.24} $ &$0.069_{-0.022}^{+0.030} $ & $-3.25_{-1.05}^{+1.24} $ & $2.25_{-0.02}^{+0.03} $ & $-0.70_{-0.24}^{+0.14} $ & $-1.58_{-0.81}^{+0.68} $\\[4pt]
050724 & $0.971_{-0.340}^{+0.352}$ & $0.232_{-0.163}^{+0.280} $ & $50.38_{-0.26}^{+0.69} $ &$0.418_{-0.137}^{+0.228} $ & $0.37_{-1.29}^{+0.44} $ & $2.25_{-0.10}^{+0.15} $ & $-0.75_{-0.44}^{+0.17} $ & $-1.11_{-0.52}^{+0.39} $\\[4pt]
050709 & $0.896_{-0.667}^{+0.449}$ & $0.319_{-0.198}^{+0.324} $ & $50.80_{-0.32}^{+0.97} $ &$0.462_{-0.371}^{+0.216} $ & $-3.83_{-0.81}^{+0.33} $ & $2.91_{-0.09}^{+0.28} $ & $-0.55_{-0.06}^{+0.03} $ & $-0.61_{-0.38}^{+0.08} $\\
\hline
\end{tabular}
\tablefoot{
\tablefoottext{a}{
For GRB~211211A, the exact posterior value is
$p=2.012^{+0.009}_{-0.007}$.
}
}
\end{table*}
\begin{table*}
\caption{Fitting results of broadband modeling of other GRB light curves given a Gaussian jet model.}
\label{tab:other}
\centering
\begin{tabular}{lcccccccc}
\hline\hline
\multirow{2}{*}{GRB} & $\theta_{\mathrm{w}}$ & $\theta_{\mathrm{obs}}$ & $\mathrm{log}~E_0$ &$\theta_{\mathrm{c}}$ & $\mathrm{log}~n_0$ & \multirow{2}{*}{$p$} & \multirow{2}{*}{$\mathrm{log}~\epsilon_{\mathrm{e}}$}& \multirow{2}{*}{$\mathrm{log}~\epsilon_{\mathrm{b}}$} \\
    &  (rad) &(rad)  &(erg) &(rad) &($\mathrm{cm^{-3}}$)&  &  & \\
\hline
211227A & $0.728_{-0.501}^{+0.436}$ &$0.257_{-0.190}^{+0.333}$& $50.97_{-0.90}^{+1.18}$ & $0.376_{-0.300}^{+0.224}$ & $-3.25_{-1.85}^{+2.32}$ & $2.22_{-0.14}^{+0.19}$&$-0.71_{-0.29}^{+0.10}$&$-3.23_{-1.19}^{+1.85}$ \\[4pt]
210726A & $0.251_{-0.118}^{+0.190}$ &$0.181_{-0.076}^{+0.119}$& $53.81_{-0.56}^{+0.55}$ & $0.071_{-0.026}^{+0.046}$ & $-1.03_{-1.43}^{+0.97}$ & $2.18_{-0.13}^{+0.18}$&$-0.72_{-0.28}^{+0.10}$&$-3.83_{-0.78}^{+2.37}$ \\[4pt]
200907B & $0.709_{-0.641}^{+0.459}$ &$0.219_{-0.200}^{+0.356}$& $50.71_{-0.41}^{+0.68}$ & $0.394_{-0.359}^{+0.261}$ & $-1.94_{-1.22}^{+1.47}$ & $4.88_{-0.21}^{+0.08}$&$-0.64_{-0.12}^{+0.05}$&$-1.06_{-1.65}^{+0.42}$ \\[4pt]
200219A & $0.752_{-0.490}^{+0.435}$ &$0.253_{-0.207}^{+0.317}$& $51.56_{-0.96}^{+1.41}$ & $0.434_{-0.354}^{+0.236}$ & $-4.11_{-1.32}^{+1.67}$ & $3.77_{-0.32}^{+0.24}$&$-0.66_{-0.14}^{+0.06}$&$-2.06_{-1.69}^{+1.17}$ \\[4pt]
181123B & $0.768_{-0.491}^{+0.430}$ &$0.281_{-0.208}^{+0.310}$& $52.23_{-0.87}^{+1.12}$ & $0.437_{-0.351}^{+0.231}$ & $-3.04_{-2.05}^{+2.20}$ & $2.28_{-0.19}^{+0.24}$&$-0.69_{-0.22}^{+0.09}$&$-3.03_{-1.34}^{+1.83}$ \\[4pt]
180727A & $0.789_{-0.458}^{+0.434}$ &$0.293_{-0.215}^{+0.297}$& $52.91_{-0.66}^{+1.09}$ & $0.446_{-0.333}^{+0.228}$ & $-4.86_{-0.82}^{+1.42}$ & $2.36_{-0.12}^{+0.13}$&$-0.65_{-0.13}^{+0.05}$&$-2.02_{-1.79}^{+1.12}$ \\[4pt]
170428A & $0.739_{-0.476}^{+0.422}$ &$0.286_{-0.205}^{+0.320}$& $51.42_{-0.99}^{+1.24}$ & $0.422_{-0.341}^{+0.244}$ & $-3.03_{-2.07}^{+2.36}$ & $2.28_{-0.19}^{+0.28}$&$-0.70_{-0.23}^{+0.09}$&$-3.10_{-1.28}^{+1.87}$ \\[4pt]
160525B & $0.728_{-0.545}^{+0.461}$ &$0.250_{-0.206}^{+0.325}$& $51.75_{-0.78}^{+1.21}$ & $0.403_{-0.338}^{+0.249}$ & $-4.73_{-0.91}^{+1.55}$ & $3.14_{-0.30}^{+0.14}$&$-0.65_{-0.12}^{+0.05}$&$-1.75_{-1.69}^{+0.96}$ \\[4pt]
160411A & $0.755_{-0.483}^{+0.433}$ &$0.270_{-0.193}^{+0.313}$& $52.11_{-0.79}^{+1.17}$ & $0.426_{-0.343}^{+0.242}$ & $-4.29_{-1.27}^{+2.19}$ & $2.62_{-0.25}^{+0.19}$&$-0.65_{-0.14}^{+0.06}$&$-1.70_{-1.76}^{+0.95}$ \\[4pt]
150728A & $0.734_{-0.493}^{+0.432}$ &$0.262_{-0.197}^{+0.320}$& $51.67_{-1.03}^{+1.30}$ & $0.418_{-0.339}^{+0.245}$ & $-4.23_{-1.25}^{+1.80}$ & $2.75_{-0.51}^{+0.83}$&$-0.66_{-0.14}^{+0.06}$&$-2.03_{-1.89}^{+1.19}$ \\[4pt]
150120A &  $0.148_{-0.104}^{+0.721}$ &$0.121_{-0.111}^{+0.448}$& $50.75_{-0.83}^{+1.36}$ & $0.064_{-0.041}^{+0.463}$ & $-1.65_{-2.78}^{+1.96}$ & $2.29_{-0.23}^{+1.21}$&$-0.68_{-0.23}^{+0.08}$&$-3.27_{-1.08}^{+1.72}$ \\[4pt]
150101B & $0.142_{-0.070}^{+0.354}$ &$0.253_{-0.123}^{+0.328}$& $52.99_{-0.66}^{+0.60}$ & $0.074_{-0.031}^{+0.332}$ & $-1.72_{-2.02}^{+1.73}$ & $2.17_{-0.11}^{+0.13}$&$-0.69_{-0.21}^{+0.09}$&$-3.90_{-0.67}^{+1.61}$ \\[4pt]
141212A & $0.449_{-0.244}^{+0.588}$ &$0.222_{-0.097}^{+0.343}$& $52.95_{-0.80}^{+0.84}$ & $0.150_{-0.089}^{+0.427}$ & $-1.87_{-2.12}^{+1.36}$ & $2.27_{-0.17}^{+0.15}$&$-0.84_{-0.45}^{+0.19}$&$-4.10_{-0.57}^{+1.08}$ \\[4pt]
140129B & $0.825_{-0.366}^{+0.441}$ &$0.300_{-0.231}^{+0.302}$& $52.76_{-0.68}^{+1.01}$ & $0.489_{-0.208}^{+0.201}$ & $-0.34_{-2.56}^{+1.01}$ & $2.42_{-0.12}^{+0.12}$&$-1.40_{-0.80}^{+0.54}$&$-3.54_{-0.75}^{+1.44}$ \\[4pt]
121226A & $0.794_{-0.421}^{+0.409}$ &$0.291_{-0.226}^{+0.308}$& $52.19_{-0.33}^{+1.01}$ & $0.465_{-0.263}^{+0.215}$ & $-4.58_{-0.82}^{+1.19}$ & $2.48_{-0.11}^{+0.07}$&$-0.66_{-0.14}^{+0.06}$&$-1.58_{-1.26}^{+0.83}$ \\[4pt]
120804A & $0.799_{-0.362}^{+0.425}$ &$0.318_{-0.237}^{+0.296}$& $52.85_{-0.18}^{+0.27}$ & $0.474_{-0.207}^{+0.206}$ & $-4.46_{-1.12}^{+1.42}$ & $2.10_{-0.02}^{+0.02}$&$-0.66_{-0.14}^{+0.06}$&$-1.78_{-0.95}^{+0.77}$ \\[4pt]
101224A & $0.214_{-0.143}^{+0.589}$ &$0.447_{-0.277}^{+0.245}$& $50.80_{-1.94}^{+2.56}$ & $0.080_{-0.040}^{+0.412}$ & $-3.04_{-2.06}^{+2.60}$ & $3.41_{-1.05}^{+1.08}$&$-0.70_{-0.30}^{+0.09}$&$-2.96_{-1.39}^{+1.86}$ \\[4pt]
100206A & $0.794_{-0.413}^{+0.425}$ &$0.288_{-0.222}^{+0.304}$& $50.69_{-1.10}^{+1.28}$ & $0.461_{-0.255}^{+0.215}$ & $-2.59_{-2.35}^{+2.19}$ & $3.30_{-0.76}^{+1.04}$&$-0.69_{-0.21}^{+0.08}$&$-2.74_{-1.53}^{+1.74}$ \\[4pt]
100117A & $0.795_{-0.399}^{+0.431}$ &$0.293_{-0.215}^{+0.306}$& $52.11_{-0.98}^{+1.15}$ & $0.464_{-0.230}^{+0.218}$ & $-3.00_{-2.05}^{+2.46}$ & $2.47_{-0.19}^{+0.17}$&$-0.71_{-0.27}^{+0.10}$&$-3.08_{-1.27}^{+1.86}$ \\[4pt]
071227 & $0.823_{-0.381}^{+0.429}$ &$0.291_{-0.222}^{+0.306}$& $51.61_{-0.83}^{+1.15}$ & $0.472_{-0.217}^{+0.210}$ & $-3.18_{-1.86}^{+2.46}$ & $2.41_{-0.19}^{+0.11}$&$-0.70_{-0.26}^{+0.09}$&$-3.22_{-1.16}^{+1.84}$ \\[4pt]
070809 & $0.823_{-0.392}^{+0.437}$ &$0.283_{-0.217}^{+0.304}$& $51.84_{-0.85}^{+1.16}$ & $0.468_{-0.215}^{+0.211}$ & $-3.23_{-1.92}^{+2.53}$ & $2.36_{-0.22}^{+0.29}$&$-0.71_{-0.27}^{+0.10}$&$-3.25_{-1.13}^{+1.80}$ \\[4pt]
070429B & $0.805_{-0.404}^{+0.436}$ &$0.290_{-0.224}^{+0.310}$& $51.72_{-0.91}^{+1.22}$ & $0.468_{-0.253}^{+0.211}$ & $-2.93_{-1.95}^{+2.28}$ & $2.58_{-0.22}^{+0.21}$&$-0.70_{-0.24}^{+0.09}$&$-3.25_{-1.15}^{+1.78}$ \\[4pt]
061210 & $0.810_{-0.382}^{+0.433}$ &$0.312_{-0.229}^{+0.305}$& $52.35_{-0.71}^{+0.87}$ & $0.468_{-0.205}^{+0.213}$ & $-3.85_{-1.52}^{+2.40}$ & $2.08_{-0.06}^{+0.06}$&$-0.71_{-0.26}^{+0.10}$&$-3.52_{-0.99}^{+1.76}$ \\[4pt]
061006 & $0.833_{-0.395}^{+0.443}$ &$0.292_{-0.219}^{+0.297}$& $52.21_{-0.78}^{+0.98}$ & $0.471_{-0.220}^{+0.208}$ & $-3.83_{-1.50}^{+2.28}$ & $2.14_{-0.06}^{+0.06}$&$-0.70_{-0.26}^{+0.10}$&$-3.52_{-1.00}^{+1.79}$ \\[4pt]
050509B & $0.726_{-0.494}^{+0.436}$ &$0.262_{-0.184}^{+0.328}$& $51.09_{-1.23}^{+1.20}$ & $0.409_{-0.335}^{+0.253}$ & $-3.35_{-1.81}^{+2.38}$ & $2.19_{-0.15}^{+0.52}$&$-0.69_{-0.24}^{+0.09}$&$-3.21_{-1.18}^{+1.90}$ \\[4pt]
\hline
\end{tabular}
\end{table*}
\clearpage
\begin{table*}
\caption{Short GRB X-ray afterglow catalog. Only the first few entries are shown here for guidance regarding its form and content.}
\label{tab:xdata} 
\centering
\begin{tabular}{lccccc}
\hline\hline
\multirow{2}{*}{GRB} &
$\delta t$ &
$\nu$ &
$F_{\mathrm{X}}$ &
$\sigma_{\mathrm{X}}$ &
\multirow{2}{*}{References} \\
 &
(hr) &
($10^{17}$ Hz) &
($\mu$Jy) &
($\mu$Jy) &
 \\
\hline
211211A      & 0.975 & 2.41 & 4.30 & 0.97 & (1), (2)\\
      & 0.992 & 2.41 & 3.26 & 0.72 & \\
      & 1.01 & 2.41 & 5.06 & 1.14 & \\
      & 1.01 & 2.41 & 7.96 & 1.78 & \\
      & 1.02 & 2.41 & 3.77 & 0.85 & \\
      & 1.04 & 2.41 & 6.12 & 1.32 & \\
\hline
\end{tabular}
\tablebib{
(1) \citet{ebp+07};
(2) \citet{ebp+09}.
}
\end{table*}

\begin{table*}
\caption{Short GRB optical/near-IR afterglow catalog. Only the first few entries are shown here for guidance regarding its form and content.}
\label{tab:otable} 
\centering
\begin{tabular}{lcccccc}
\hline\hline
\multirow{2}{*}{GRB} &
$\delta t$ &
\multirow{2}{*}{Telescope} &
\multirow{2}{*}{Band} &
$F_{\nu}$ &
$\sigma_{\nu}$ &
References \\
 &
(hr) &
 &
 &
($\mu$Jy) &
($\mu$Jy) &
 \\
\hline
211211A  & 0.961 & Swift/UVOT & b & 50.1 & 11.8 & (1)\\
    & 1.20 &  & v& 78.0 & 23.9 & \\
      & 1.20 &  & uvm2& 53.0 & 8.9 & \\
211023B       & 7.44 & CAHA/CAFOS & $\mathrm{i^{'}}$  & 4.72 & 0.45 & (2)\\
      & 41.9 & Gemini/GMOS-N & $\mathrm{i}$ & 2.18 & 0.21 &(3) \\
        & 14.9 & OAN-SPM/RATIR&i & $<5.48$ &   &(4) \\
        & 0.347 &Swift/UVOT & uvw1 & $<19.9$ &   & (5)\\
\hline
\end{tabular}
\tablebib{
(1) \citet{rgl+22};
(2) \citet{kdt+21};
(3) \citet{rpf+21};
(4) \citet{wbk+21};
(5) \citet{kds+21}.
}
\end{table*}

\begin{table*}
\caption{Short GRB radio afterglow catalog. Only the first few entries are shown here for guidance regarding its form and content.}
\label{tab:rtable} 
\centering
\begin{tabular}{lcccccc}
\hline\hline
\multirow{2}{*}{GRB} &
$\delta t$ &
\multirow{2}{*}{Facility} &
Mean frequency &
$F_{\nu}$ &
$\sigma_{\nu}$ &
\multirow{2}{*}{References} \\
 &
(hr) &
 &
(GHz) &
($\mu$Jy) &
($\mu$Jy) &
 \\
\hline
211023B        & 87.5 & VLA & 6.05 & $<13.0$ &   &(1) \\
210726A       & 281 &MeerKAT &1.28 & 19.0 & 6.0 &(2)  \\
      & 569 & &1.28 & 24.0 & 6.0 &  \\
      & 1.00$\times 10^{3}$ & &1.28 & 21.0 & 6.0 &\\
        & 43.3 & &1.28 & $<24.0$ &   & \\
        & 137 & &1.28 & $<27.0$ &   & \\
        & 1.48$\times 10^{3}$ & &1.28 & $<21.0$ &   & \\
\hline
\end{tabular}
\tablebib{
(1) \citet{sfr+21};
(2) \citet{srl+24}.
}
\end{table*}
\twocolumn
\newpage

\section{Results from the full sample}
\label{appendix:full}

\begin{figure}
\centering
\includegraphics[width=0.99\columnwidth]{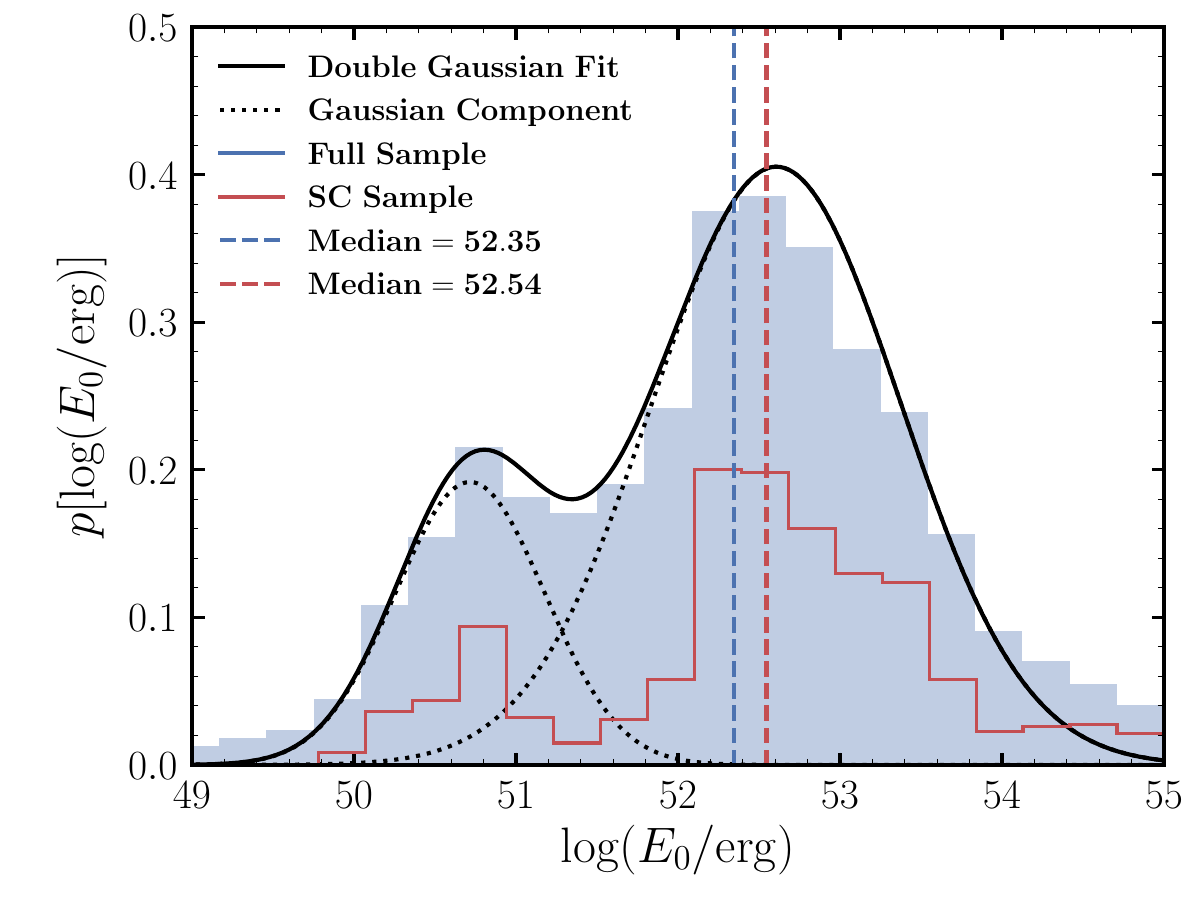}
\includegraphics[width=0.99\columnwidth]{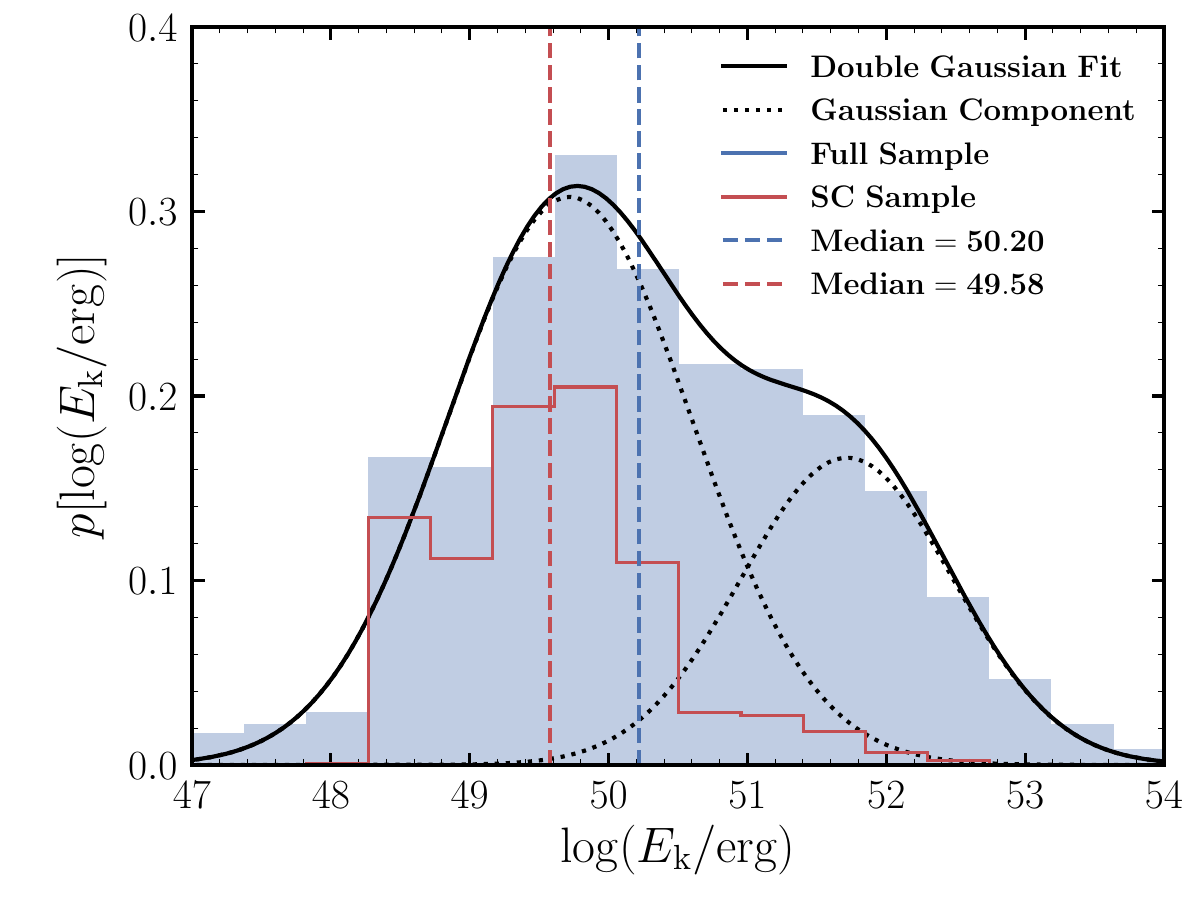}
\includegraphics[width=0.99\columnwidth]{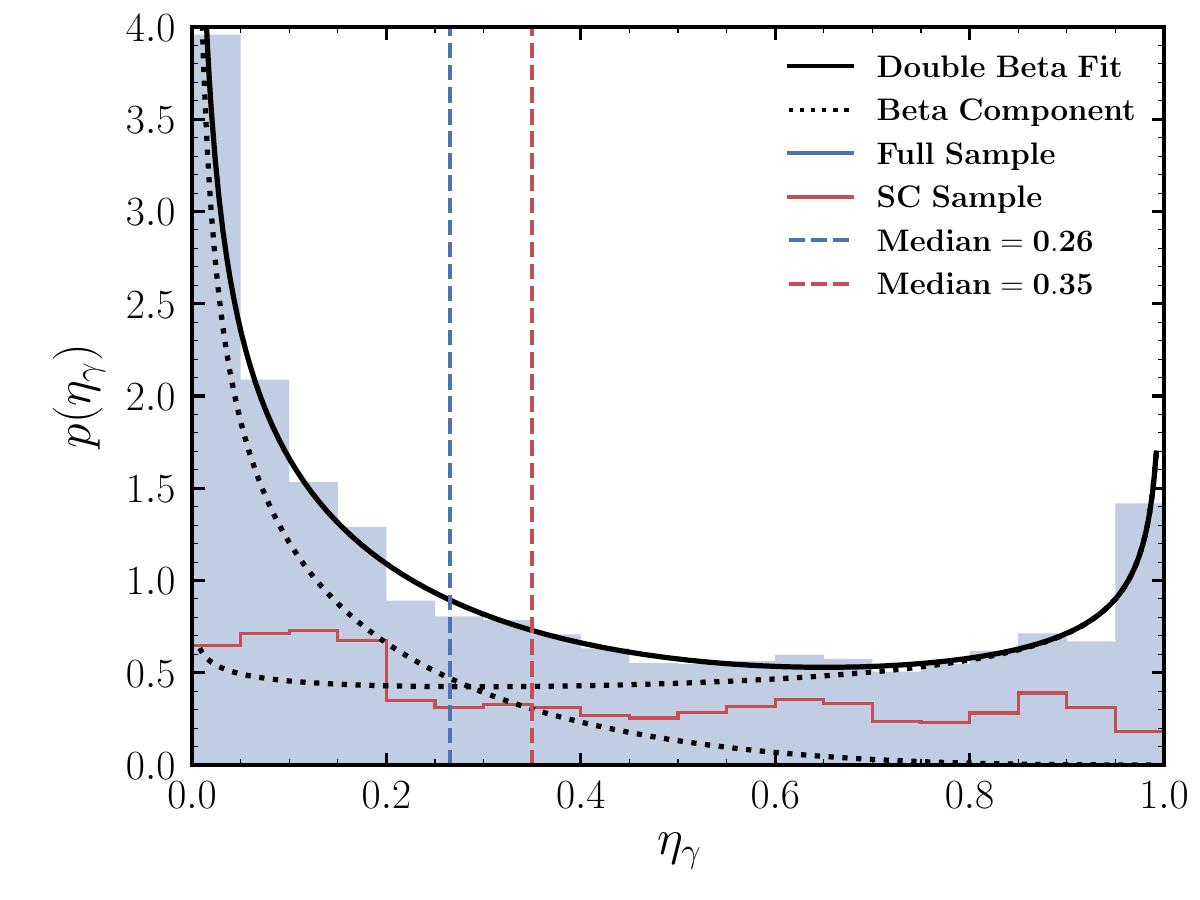}
\caption{
Comparison of the inferred distributions of the on-axis energy $E_0$ (top panel), the beaming-corrected kinetic energy $E_\mathrm{k}$ (middle panel), and the radiative efficiency $\eta_\gamma$ (bottom panel) for the full and SC samples. The blue shaded histograms represent the full sample of 40 sGRBs, while the red step histograms correspond to the 15 SC bursts.
The black solid curves show the double-Gaussian fits for $E_0$ and $E_\mathrm{k}$ and the double-Beta fit for $\eta_\gamma$ obtained from the full sample, while the black dotted curves represent their individual components. 
The vertical dashed lines indicate the median values of the corresponding distributions, with the same color coding as the histograms.
}
\label{fig:full_E0}
\end{figure}

For comparison, we extend the statistical analysis to the full sample of 40 GRBs, including the 15 SC and 25 non-SC bursts. Figure~\ref{fig:full_E0} compares the distributions of the on-axis energy $E_0$, the beaming-corrected kinetic energy $E_\mathrm{k}$, and the radiative efficiency $\eta_\gamma$ between the SC and full samples in three panels. This comparison allows us to examine how the inclusion of the less well-constrained bursts affects the inferred distributions.

For the full sample case, the median value is $\log(E_0/\mathrm{erg})=52.35$ , with a $68\%$ credible interval of [50.87, 53.37]. This is close to the median value of 52.54 obtained for the SC sample, and the two samples exhibit similar distributions (see the top panel of Figure~\ref{fig:full_E0}). This agreement indicates that the inferred $E_0$ distribution is relatively insensitive to the inclusion of the non-SC bursts, consistent with the fact that $E_0$ is generally well constrained for individual bursts. The full sample distribution is well described by the double-Gaussian function given in Equation~(\ref{con:E0}), with best-fitting parameters $A_1=0.22$, $\mu_1=50.72$, $\sigma_1=0.46$, $\mu_2=52.61$, and $\sigma_2=0.77$.

However, the situation is different for $E_\mathrm{k}$ and $\eta_\gamma$. For the full sample, the median values are $\log(E_\mathrm{k}/\mathrm{erg})=50.20$ and $\eta_\gamma=0.26$, with $68\%$ credible intervals of [49.00, 51.74] and [0.03, 0.76], respectively. The corresponding distributions are fitted using double-Gaussian and double-Beta functions, as defined in Equations (\ref{con:E0}) and (\ref{con:beta}), respectively. The best-fitting parameters for the double-Gaussian model are $(A_1,\mu_1,\sigma_1,\mu_2,\sigma_2)=(0.68,49.72,0.88,51.72,0.76)$, while those for the double-Beta model are $(\alpha_1,\beta_1,\alpha_2,\beta_2,A_1)=(0.58,3.60,0.87,0.65,0.46)$. Both distributions differ noticeably from those of the SC sample, as shown in the middle and bottom panels of Figure~\ref{fig:full_E0}. These differences are likely related to the weaker constraints on the jet geometry for the non-SC bursts. For these sources, the sparse afterglow data provide only limited constraints on the absolute angular scale of the jet. Although the empirical prior constrains the ratio $\theta_{\mathrm{w}}/\theta_{\mathrm{c}}$, relatively large values of $\theta_{\mathrm{c}}$ remain allowed, with correspondingly large $\theta_{\mathrm{w}}$. In these poorly constrained cases, larger $\theta_{\mathrm{c}}$ can also be accommodated by larger viewing angles, since different combinations of $\theta_{\mathrm{c}}$ and $\theta_{\mathrm{obs}}$ can reproduce similarly shaped light curves within the available data. As shown by comparing Figs.~\ref{fig:well} and \ref{fig:others}, the non-SC bursts generally have fewer afterglow observations or lack sufficiently late-time coverage, particularly around the expected jet-break epoch. Such observations are especially important for constraining the characteristic angular scale of the jet and reducing degeneracies among the geometrical parameters. The relatively weak constraints on the jet geometry of these bursts are therefore understandable. For a Gaussian jet, a larger $\theta_{\mathrm{c}}$ corresponds to a broader energetic core and consequently increases the beaming-corrected kinetic energy obtained by integrating the angular energy profile. The enhanced high-energy tail in the full sample distribution is therefore driven by the weakly constrained jet geometry of the non-SC bursts. This further highlights the importance of well-sampled, particularly late-time, afterglow observations for obtaining reliable constraints on GRB jet geometry and energetics.

\begin{figure}
\centering
\includegraphics[width=0.99\columnwidth]{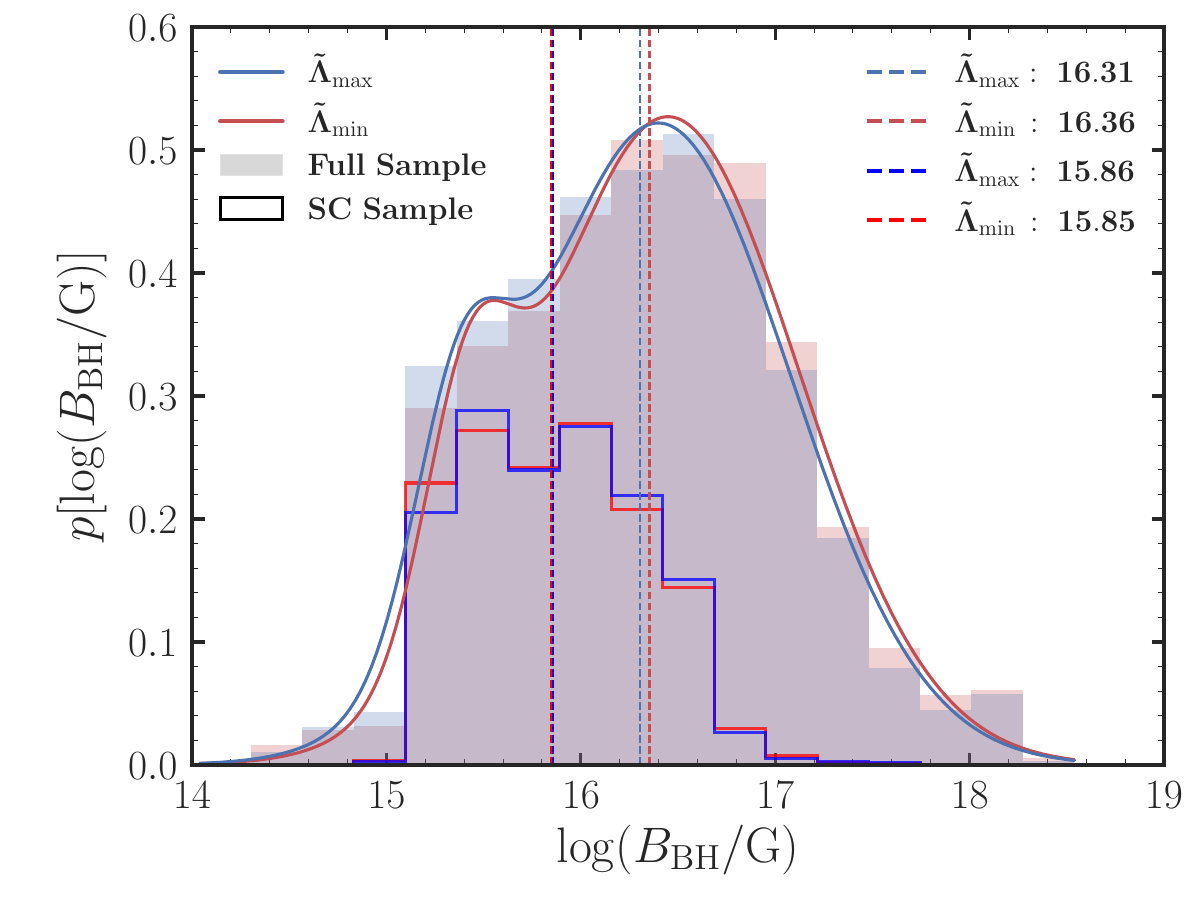}
\includegraphics[width=0.99\columnwidth]{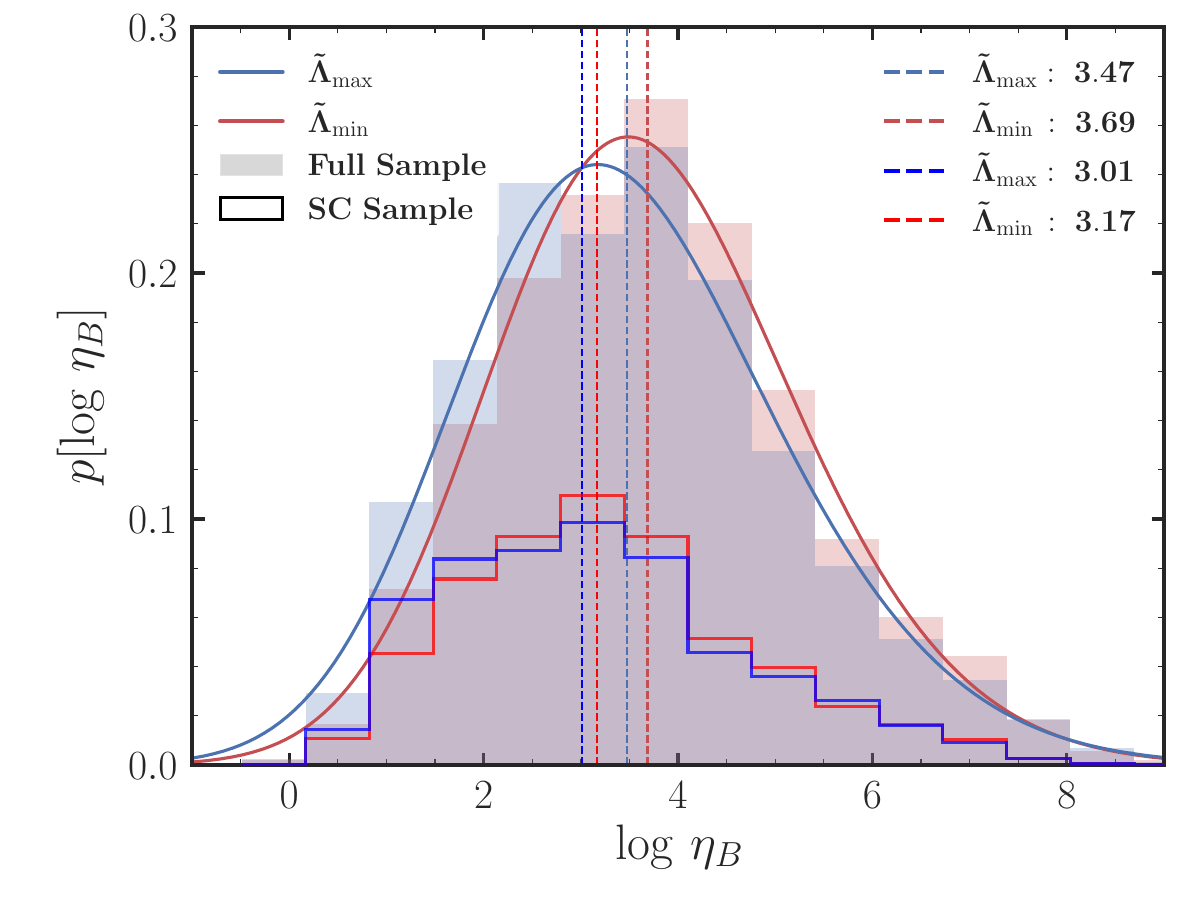}
%
%
\caption{
Comparison of the inferred distributions of the BH magnetic field strength $B_\mathrm{BH}$ (top panel) and the magnetic field amplification factor $\eta_B$ (bottom panel) for the full and SC samples. The blue and red histograms correspond to the $\tilde{\Lambda}_{\mathrm{max}}$ and $\tilde{\Lambda}_{\mathrm{min}}$ assumptions, respectively. The shaded histograms represent the full sample of 40 sGRBs, while the unfilled step histograms represent the 15 SC bursts. The solid curves show the double-Gaussian fits for $B_\mathrm{BH}$ and the skew-normal fits for $\log\eta_B$ obtained from the full sample. The vertical dashed lines indicate the corresponding median values, with lighter and darker lines representing the full and SC samples, respectively.
}
\label{fig:full_B}
\end{figure}

Figure~\ref{fig:full_B} compares the inferred distributions of the BH magnetic field strength $B_\mathrm{BH}$ (top panel) and the magnetic field amplification factor $\eta_B$ (bottom panel) between the SC and full samples under the two $\tilde{\Lambda}$ assumptions. Similar to $E_\mathrm{k}$ and $\eta_\gamma$, the full sample distributions are shifted toward higher values compared to those of the SC sample, while substantial overlap remains between them. For $\tilde{\Lambda}_{\mathrm{max}}$, the full sample yields median values of $\log(B_\mathrm{BH}/\mathrm{G})=16.31$ and $\eta_B=2859$, with corresponding $68\%$ credible intervals of [15.51, 17.00] and [80, 126109], respectively. The distributions of $\log(B_\mathrm{BH}/\mathrm{G})$ and $\log\eta_B$ are fitted using double-Gaussian and SG functions, as defined in Equations~(\ref{con:E0}) and (\ref{con:skewed}), respectively. The best-fitting parameters are $(A_1,\mu_1,\sigma_1,\mu_2,\sigma_2)=(0.11,15.37,0.24,16.40,0.68)$ for the double-Gaussian model and $(\alpha,\mu,\sigma)=(1.64,1.93,2.30)$ for the SG model. For $\tilde{\Lambda}_{\mathrm{min}}$, the median values are 16.36 and 4853, with $68\%$ credible intervals of [15.54, 17.04] and [158, 196938], respectively. The best-fitting parameters are $(A_1,\mu_1,\sigma_1,\mu_2,\sigma_2)=(0.11,15.42,0.23,16.45,0.67)$ for the double-Gaussian model and $(\alpha,\mu,\sigma)=(1.45,2.34,2.11)$ for the SG model. Both the inferred magnetic field strengths and amplification factors have higher median values than those obtained for the SC sample. However, the characteristic magnetic field strength remains of order $10^{16}$ G, indicating that our main conclusion remains robust to the sample selection.

\end{appendix}
\end{document}